\pdfoutput=1
\documentclass[aip,jcp,amsmath,amssymb,preprint,]{revtex4-1}

\usepackage{bm}
\usepackage{dcolumn}

\usepackage{hyperref}
\usepackage{graphicx}
\usepackage{amsfonts,amsmath,amssymb}
\usepackage{color}
\usepackage{multirow}
\usepackage[normalem]{ulem}

\newcommand{\inv}{\ensuremath{{^{\raisebox{.2ex}{$\scriptscriptstyle-1$}}}}}

\newcolumntype{d}{D{.}{.}{3.1}}
\newcommand{\mcone}[1]{\multicolumn{1}{c}{#1}}

\begin{document}

\title{On the accuracy of the MB-pol many-body potential for water:
Interaction energies, vibrational frequencies, and classical thermodynamic and dynamical properties from clusters to liquid water and ice.}

\author{Sandeep K. Reddy}
\affiliation{Department of Chemistry and Biochemistry, University of California, 
San Diego, La Jolla, California 92093, United States}

\author{Shelby C. Straight}
\affiliation{Department of Chemistry and Biochemistry, University of California, 
San Diego, La Jolla, California 92093, United States}

\author{Pushp Bajaj}
\affiliation{Department of Chemistry and Biochemistry, University of California, 
San Diego, La Jolla, California 92093, United States}

\author{C. Huy Pham}
\affiliation{Department of Chemistry and Biochemistry, University of California, 
San Diego, La Jolla, California 92093, United States}

\author{Marc Riera}
\affiliation{Department of Chemistry and Biochemistry, University of California, 
San Diego, La Jolla, California 92093, United States}

\author{Daniel R. Moberg}
\affiliation{Department of Chemistry and Biochemistry, University of California, 
San Diego, La Jolla, California 92093, United States}

\author{Miguel A. Morales} 
\affiliation{Lawrence Livermore National Laboratory, 7000 East Avenue, Livermore, 
California 94550, U.S.A.}

\author{Chris Knight} 
\affiliation{Leadership Computing Facility, Argonne National Laboratory, 
9700 South Cass Avenue, Argonne, Illinois 60439, United States}

\author{Andreas W. G\"{o}tz}
\affiliation{San Diego Supercomputer Center, University of California, San Diego, 
La Jolla, California 92093, United States}

\author{Francesco Paesani}
\affiliation{Department of Chemistry and Biochemistry, University of California, 
San Diego, La Jolla, California 92093, United States}

\date{\today}

\begin{abstract}

The MB-pol many-body potential has recently emerged
as an accurate molecular model for water simulations from
the gas to the condensed phase.
In this study, the accuracy of MB-pol is systematically assessed across the three phases
of water through extensive comparisons with experimental data and
high-level \textit{ab initio} calculations.  
Individual many-body contributions to the interaction energies as well as vibrational spectra
of water clusters calculated with MB-pol are in excellent agreement with reference data obtained 
at the coupled cluster level. Several structural, thermodynamic, and dynamical
properties of the liquid phase at atmospheric pressure are investigated through classical molecular
dynamics simulations as a function of temperature. 
The structural properties of the liquid phase are in nearly quantitative 
agreement with X-ray diffraction data available over the temperature range from 268 to 368~K. 
The analysis of other thermodynamic and dynamical quantities emphasizes the 
importance of explicitly including nuclear quantum effects in the simulations, 
especially at low temperature, for a physically correct description of the properties of liquid water. 
Furthermore, both densities and lattice energies of several ice
phases are also correctly reproduced by MB-pol. 
Following a recent study of DFT models for water, a score is assigned to each computed
property, which demonstrates the high and, in many respects, unprecedented 
accuracy of MB-pol in representing all three phases of water.

\end{abstract}

\maketitle

\bibliographystyle{apsrev}

\section{Introduction} 
Fast, reliable, and accurate modeling of structural, physical, and chemical
properties of water across all media - gas, liquid, interface, confined and
solid - and at different thermodynamic conditions is a long-standing challenge.
This is largely because the quality of any theoretical prediction depends
heavily on the underlying intermolecular potential energy surface (PES)
utilized.\cite{FrancescoChemRev2016,CCpol6,
CCpol3,VegaPCCP2011} 
While in principle, \textit{ab initio} molecular dynamics (AIMD) simulations
carried out with correlated electronic structure methods can provide
a correct description of water across all phases, these simulations
are currently unfeasible due to the prohibitive associated computational cost.
Despite recent progress in the development and implementation of 
simulation approaches based on M\"{o}ller-Plesset perturbation theory,
density functional theory (DFT) still remains the most common approach
used in \textit{ab initio} simulations of water.\cite{MarxHutterBook,ParrinelloPRL1985,Voth2002,Payne1992}
However, existing functionals have been shown to remain limited in
the accuracy and predictive ability with which they can represent the properties 
of water in different phases.\cite{AngelosJCP2016}

On the other hand, empirical water models based on molecular mechanics
(commonly referred to as force fields) have been used extensively in computer
simulations to investigate the properties of water under different
thermodynamic conditions. In these models, the force field parameters are
usually tuned to reproduce a (limited) set of experimental
properties.\cite{Bertrand2002,SadusFPE2016,tip4p_Vega2005,
tip4p_Gordon2004,JorgensenTIP4P1983,JorgensenJCP2000,Berendsen1987,
VegaFarad2009,VegaPCCP2011} The most common empirical models are pairwise
additive, assume the water molecules to be rigid, and use fixed point charges
to describe the electrostatic interactions. Despite their simplicity, empirical
models have had great success in reproducing, at least to some extent,
structural, thermodynamic, and transport properties of liquid water over a broad
range of temperatures and pressures.\cite{tip4p_Vega2005,
tip4p_Gordon2004,Bertrand2002,JorgensenTIP4P1983,JorgensenJCP2000,Berendsen1987,
VegaFarad2009,VegaPCCP2011} Examples under this category include
RWK,\cite{KleinCP1982} SPC,\cite{Berendsen1981} SPC/E,\cite{Berendsen1987}
TIP4P,\cite{JorgensenTIP4P1983} TIP4P-2005,\cite{tip4p_Vega2005}
TIP4P-Ew,\cite{tip4p_Gordon2004} and TIP5P.\cite{JorgensenJCP2000} The reader
is referred to Refs.~\citenum{VegaPCCP2011},~\citenum{Bertrand2002} and
\citenum{SadusFPE2016} as well as the original studies for complete details of
rigid water models.  Flexible versions were also developed for some of these
water models to investigate vibrational dynamics and energy transfer in the
liquid phase.\cite{ManolopoulosJCP2009,Francesco2006} To go beyond the pairwise
approximation, several water models have been developed which include either
empirical three-body terms (e.g., E3B\cite{Skinner2008,Skinner2011,Skinner2015}) or implicit many-body
effects through classical polarization (e.g., BK3,\cite{bk3}
SWM4-DP,\cite{swm4dp} SWM4-NDP,\cite{swm4ndp} SWM6,\cite{swm6}
COS,\cite{cos1,cos2,bk3} TTMx-F,\cite{TTM2,TTM3,TTM4}
AMOEBA,\cite{amoeba2003,AMOEBA2,iamoeba,AMOEBA2015} GEM*,\cite{gems} and
POLY2VS\cite{Yoshitaka2011}).  The interested reader is referred to Ref.
\citenum{FrancescoChemRev2016} for a systematic overview of recent water models.

While water models parameterized to reproduce experimental data have been 
instrumental in gaining qualitative insights into the behavior of liquid water, the
lack of chemical accuracy results in limited predictive power across the entire
phase diagram.
Over the years, this has stimulated the development of analytical potential energy functions,
originally introduced by Clementi and co-workers in the 1970s,
which aim at representing the multidimensional potential energy surface (PES)
associated with a system containing N water molecules through 
a rigorous representation of the many-body expansion (MBE) of the interaction energy,\cite{hankins_d_1970}
\begin{equation} 
\label{eq:mbexpansion} 
E_{\text{N}} = \sum_{i=1}^N
V^{\text{1B}}(i) + \sum_{i<j}^NV^{\text{2B}}(i,j) + \sum_{i<j<k}^N
V^{\text{3B}}(i,j,k) + \dots + V^{\text{NB}}(1,\dots,N) , 
\end{equation}
where $V^{\text{1B}}$ is the one-body (1B) contribution corresponding to 
the deformation energy and the $V^{\text{nB}}$ are
the n-body (nB) interaction energies defined as
\begin{equation} \label{eq:nbody} \begin{split} V^{\text{nB}} = \quad &
E_n(1,\dots,n) - \sum_iV^{\text{1B}}(i) - \sum_{i<j}V^{\text{2B}}(i,j) - \dots
\\ & - \sum_{i<j<\dots<n-1}V^{\text{(n-1)B}}(i,j,\dots,(n-1)) . \end{split}
\end{equation}
Depending on the specific functional form adopted and
the extent of the training sets used to fit the individual terms of the MBE, 
many-body potentials can approach the same level of
accuracy as high-level correlated electronic structure calculations at a fraction of
the computational cost.\cite{Francesco1JCTC2013}
The most notable many-body potential energy functions 
are CC-pol,\cite{CCpol1,CCpol2,CCpol3,CCpol4,CCpol5,CCpol6}
WHBB,\cite{WHBB1,WHBB2,WHBB3,WHBB4,WHBB5}
HBB2-pol,\cite{FrancescoJPCL2012,Francesco1JCTC2013,babin_v_2013}
and MB-pol.\cite{FrancescoJCTC2013,
FrancescoJCTC2014,Francesco1JCTC2014,FrancescoJCTC2015}
To date, MB-pol (and its precursor HBB2-pol) is the only many-body potential that has been 
consistently employed in molecular simulations, with explicit inclusion
of nuclear quantum effects (NQE), which
correctly reproduce the properties of water
from the gas to the liquid and solid phases.
\cite{FrancescoJPCL2012,babin_v_2013,
FrancescoJCTC2013,FrancescoJCTC2014,Francesco1JCTC2014,FrancescoJCTC2015,
FrancescoJCP2015,Francesco1JCP2015,FrancescoJACS2016,FrancescoJPCB2016}

The purpose of this study is to systematically assess the accuracy of MB-pol
in predicting structural, thermodynamic, dynamical, and spectroscopic 
properties of water across all phases as well as 
to provide a metric by which these properties
can be compared to experiment. 
The article is organized as follows. Section II describes the technical details
of all calculations and simulations employed in this study. 
Section III reports the results on several physical
properties of water in the three phases: gas, liquid, and solid. The overall
performance of MB-pol is then discussed by assigning a score to each computed
property. The last section summarizes the results and provides an outlook of future
applications of MB-pol.

\section{Methodology and Computational details} 
\label{section:methods} 
As discussed in detail in previous studies,\cite{Francesco1JCTC2013,FrancescoJCTC2013,
FrancescoJCTC2014,Francesco1JCTC2014,FrancescoJCTC2015}
the MB-pol potential is built upon the MBE of Eq.~\ref{eq:mbexpansion} and includes
explicit terms that describe 1B, 2B, and 3B terms, along with classical
N-body polarization to account for all higher-body contributions to the interaction energy. 
The polarization term is represented by a slightly modified version of the Thole-type model as
introduced in the TTM4-F model.\cite{TTM4} The 1B term is represented by the
spectroscopically accurate water monomer PES developed by Partridge and
Schwenke.\cite{SchwenkeJCP1997} The 2B term is further divided into long-range
and short-range interactions and is described using classical electrostatics,
induction, and dispersion forces, which dominate the long-range part,
supplemented by a set of multivariable polynomials in the short-range, to
capture the more complex quantum mechanical effects arising from the overlap of
monomer electron densities. Along the same lines, the 3B term is composed of 3B
induction and a multi-dimensional function to accurately describe both
long-range and short-range interactions. The polynomial functions used to
describe the short-range 2B and 3B interactions are generated in such a way
that they retain permutational invariance with respect to the hydrogen atoms within
the same water molecule as well as to whole water molecules within all possible
dimers and trimers.
The permutationally invariant polynomials are trained to a large set of highly
accurate correlated coupled-cluster energies via a supervised machine learning
approach. Specific details about the functional form, training sets, and
training algorithms can be found in the original
studies.\cite{FrancescoJPCL2012,Francesco1JCTC2013,FrancescoJCTC2013,
FrancescoJCTC2014,Francesco1JCTC2014,FrancescoJCTC2015} MB-pol is publicly available 
as an external plugin\cite{MBpol_openmm} for the OpenMM toolkit for molecular
simulations\cite{PandeJCTC2013} and has recently been interfaced\cite{MBpol_ipi} to the i-PI
wrapper.\cite{CeriottiCPC2014}

All electronic structure calculations of water clusters presented in the next
sections were carried out using MOLPRO.\cite{MOLPRO_brief} The reference
interaction energies for (H$_2$O)$_\text{n}$ clusters, with n = 4$-$6, optimized at
the RI-MP2 and MP2 levels of theory in Refs. \citenum{temelso_b_2011} and
\citenum{bates_dm_2009}, were obtained using the MBE of the interaction
energy\cite{hankins_d_1970} as described in Ref. \citenum{gora_u_2011}, with
individual MB contributions calculated with the CCSD(T)/CCSD(T)-F12b method in
the complete basis set (CBS) limit.\cite{hill_jg_2009,neese_f_2011} The
2B interaction energies were computed at the CCSD(T) level of theory by extrapolating
the values obtained with aug-cc-pVTZ and aug-cc-pVQZ basis sets supplemented
with an additional set of (3s, 3p, 2d, 1f) midbond functions, with exponents
equal to (0.9, 0.3, 0.1) for s and p orbitals, (0.6, 0.2) for d orbitals, and 0.3 for f orbitals,
placed at the center of mass of each dimer
configuration.\cite{DunningJCP1989,Kendall1992,Tao1992} 
The following two-point formula\cite{halkier_a_1999, halkier_a_1999b} was used 
to extrapolate the interaction energies to the CBS limit: 
\begin{equation}
\text{V}_{\text{2B}}^{\text{X}} = \text{V}_{\text{2B}}^{\text{CBS}} - \frac{\text{A}} {\text{X}^{\text{3}}}
\end{equation}
with cardinal numbers X = 3 and 4, accordingly. The Hartree--Fock energy was not extrapolated 
separately since it was close to the CBS limit for either value of X. 
The 3B interaction energies
were calculated at the CCSD(T) level of theory using the aug-cc-pVTZ basis
set supplemented with the same set of midbond functions introduced above, which were placed 
at the center of mass of each trimer configuration.  All higher ($>$3B) contributions were
computed with the CCSD(T)-F12 method using the VTZ-F12 basis
sets.\cite{WernerJCP2007,WernerJCP2008,WernerJCP2009} This method yields
results close to the CBS values at lower computational cost than direct CCSD(T) calculations
with large basis sets.\cite{tew_dp_2007, bischoff_fa_2009} All
3B and higher-body energies were corrected for the basis set superposition
error (BSSE) using the counterpoise method.\cite{boys_sf_1970}

The MB-pol vibrational frequencies of the water clusters were calculated within the harmonic
approximation from the diagonalization of the mass-weighted Hessian matrix.
Each cluster structure was first energy minimized until the norm of the force vector reached 
a value smaller than 10$^{-8}$ kcal mol\inv~\AA$^{-1}$.  
The absence of any imaginary frequencies indicates
that all structures reported in Section~\ref{subsection:vib_freq} 
correspond to either a local or a global minimum.

All molecular dynamics (MD) simulations of liquid water presented in the next sections
were carried out at the classical level using in-house software based on 
the DL\_POLY\_2 simulation package,\cite{DL_POLY1996} which was
modified to include the MB-pol potential.\cite{Francesco1JCTC2014}
Unless otherwise stated, the systems consisted of 256 water molecules placed 
in a cubic simulation box.  The velocity-Verlet algorithm was used to integrate Newton's 
equations of motion with a timestep of 0.2 fs. 
All thermodynamic properties except
the surface tension were calculated from simulations carried out in the
isobaric-isothermal (NPT) ensemble.
The temperature and pressure were kept constant using the
Nos{\'{e}}-Hoover thermostat and barostat,
respectively.\cite{Nose1984,Nose1992,Hoover1985,Ciccotti1993} 
The short-range interactions were evaluated within a distance cutoff of 9 \AA. 
Short-range electrostatic interactions were computed in real space using
Coulomb's law while the long range interactions were calculated in reciprocal
space using the Ewald summation technique.  A high precision was used (10$^{-8}$) to
generate the best Ewald vectors and the Ewald convergence parameter for MD
simulations. The MD simulations were run in the NPT ensemble at atmospheric
pressure (P = 1 atm) for twelve temperatures between 248~K and 368~K.  
The trajectory lengths at different temperatures are listed in
Table~XL of the Supplementary Material.
During the production run, the positions and dipole
moments were collected for analysis.  To compute the isothermal compressibility,
additional simulations were run for 4.5 ns at each temperature for the following
pressures: -1.5, -1.0, -0.5, 0.5, 1.0, and 1.5 katm.  

To calculate the surface tension of liquid water, a slab geometry was prepared from a
fully equilibrated bulk simulation of 511 
water molecules in a cubic box by
expanding the z-axis of the box to 100 \AA. After preparation of the initial
slab geometry, equilibration trajectories of 500 ps were simulated 
in the isochoric-isothermal (NVT) ensemble by employing
periodic boundary conditions and Ewald summation in all three dimensions. 
Following the equilibration, a 1.6 ns trajectory was generated at each temperature
for analysis.
The self-diffusion coefficient was calculated, at each temperature, by
averaging over thirty independent 100 ps long trajectories carried out
in the microcanonical (NVE) ensemble.
The initial configurations for the NVE trajectories were obtained from
thirty 10 ps long NVT trajectories started from configurations
extracted at intervals of 50 ps from an equilibrated NPT trajectory.

For each ice phase, the system size was chosen such that the 
edges of the simulation box were always separated by more than 18 \AA. 
The ice phases considered in this study are ice I$_h$, the
proton disordered phase at ambient conditions, and several proton ordered phases
(II, VIII, IX, XIII, XIV, and XV).  The initial structures for the proton
ordered phases were taken from Ref.~\citenum{PhysRevLett.107.185701}, while for
ice I$_\textrm{h}$, 13 independent configurations were generated by minimizing the net dipole moment
following the algorithm proposed in Ref.~\citenum{buch1998simulations}.
One additional configuration for ice I$_\textrm{h}$ taken from Ref.
\citenum{PhysRevLett.107.185701} was also included in the calculations.
All configurations satisfy the Bernal-Fowler rules.\cite{bernal1933theory} 
The densities of the different ice phases were calculated by averaging over 100 ps 
of NPT simulations.

\section{Results and Discussion}

\subsection{Water clusters}
\label{subsection:clusters}
\subsubsection{Many-body energy decomposition} 
\label{subsection:mbe}

\begin{figure*}[b]
 \begin{center}
  \includegraphics[width=1.0\textwidth]{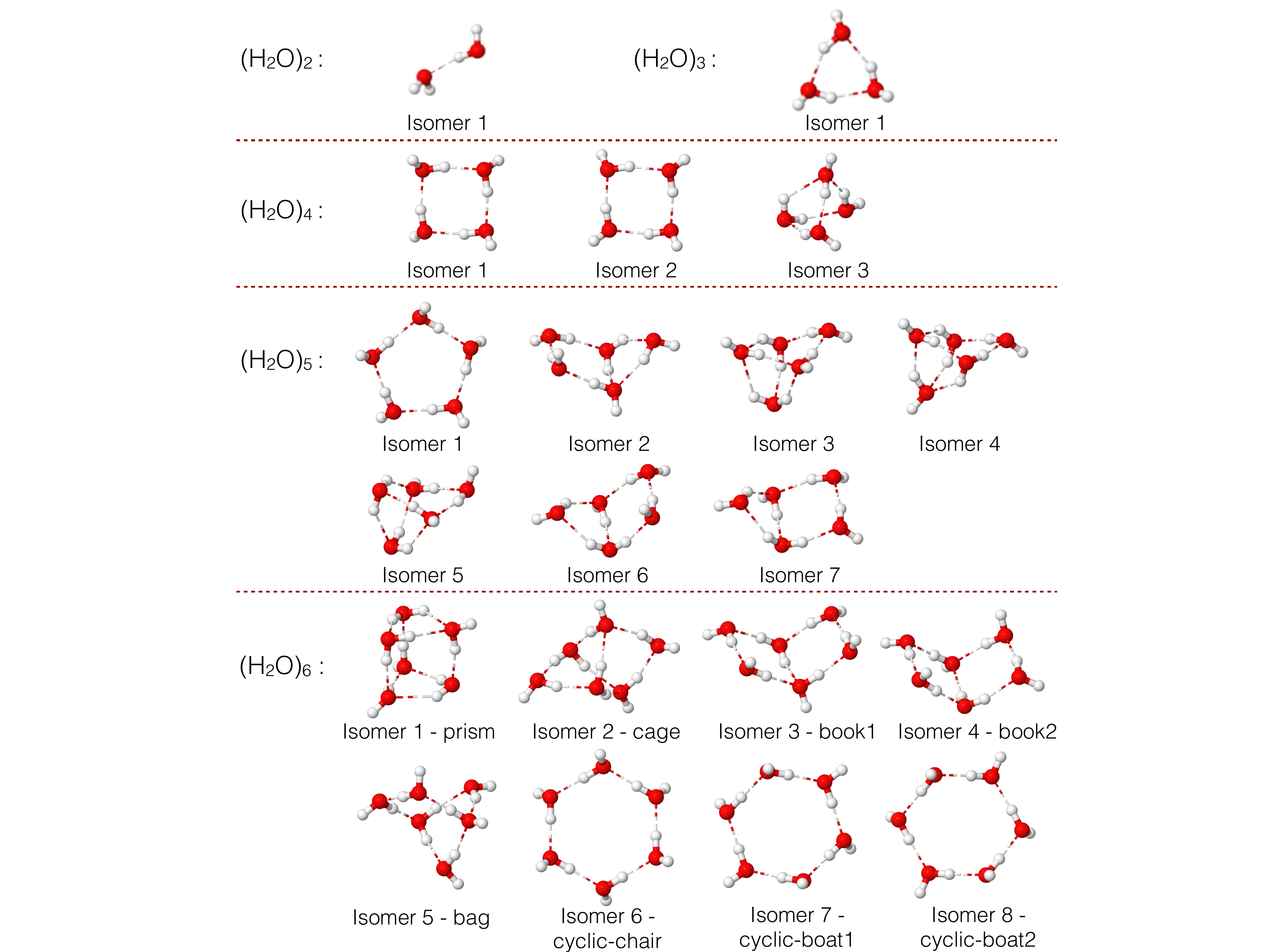} 
 \end{center} 
  \caption{(H$_2$O)$_\text{n}$ clusters used for the analysis of the many-body decomposition
  of the total interaction energies and harmonic frequencies.}
  \label{f:water-clusters}
\end{figure*}

\begin{figure}[t]
\begin{center}
\includegraphics[width=1.0\textwidth]{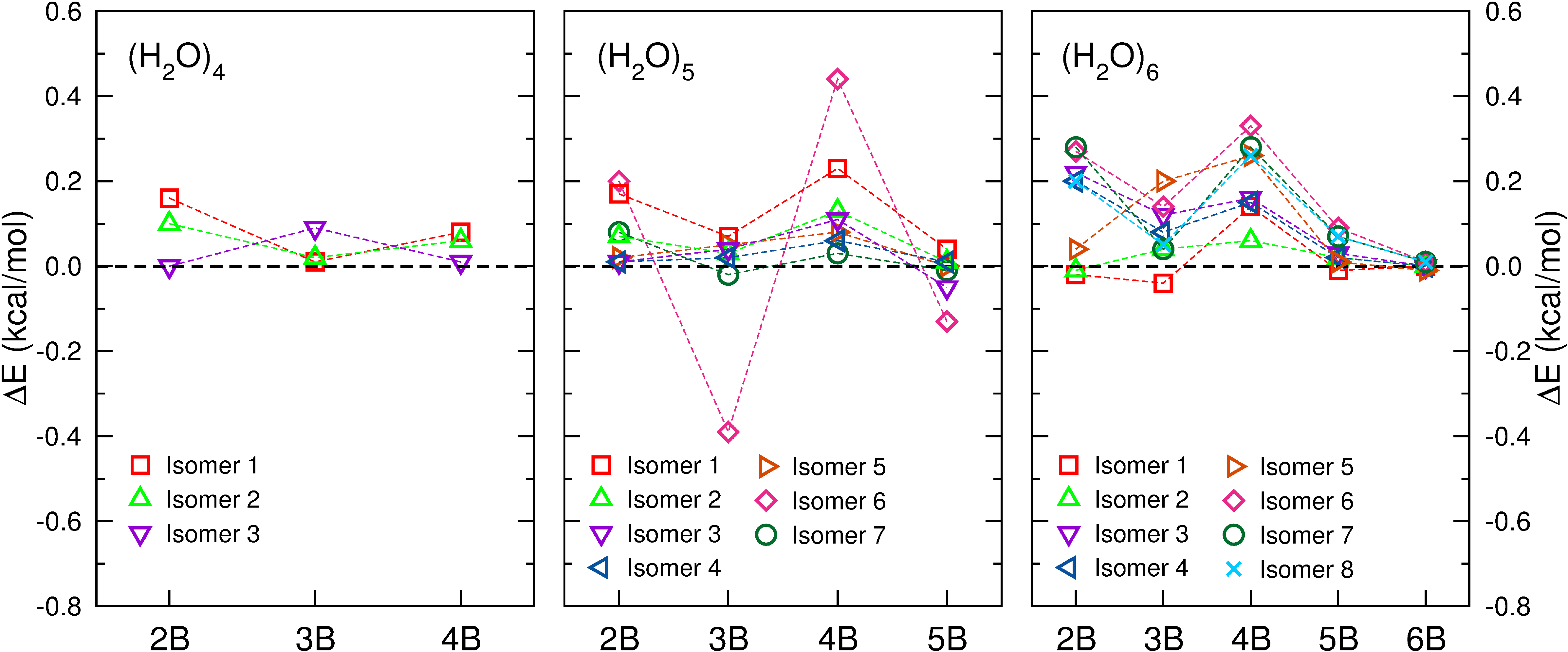}
\end{center} 
\caption{Errors in the MB decomposition of the interaction energy,
relative to the corresponding values calculated with the CCSD(T)/CCSD(T)-F12
method in the CBS limit, for the (H$_2$O)$_\text{n}$ (n=4, 5, 6) clusters.
}
\label{fig:clustermbdecomposition} 
\end{figure}

\begin{figure}
\begin{center}
\includegraphics[width=1.0\textwidth]{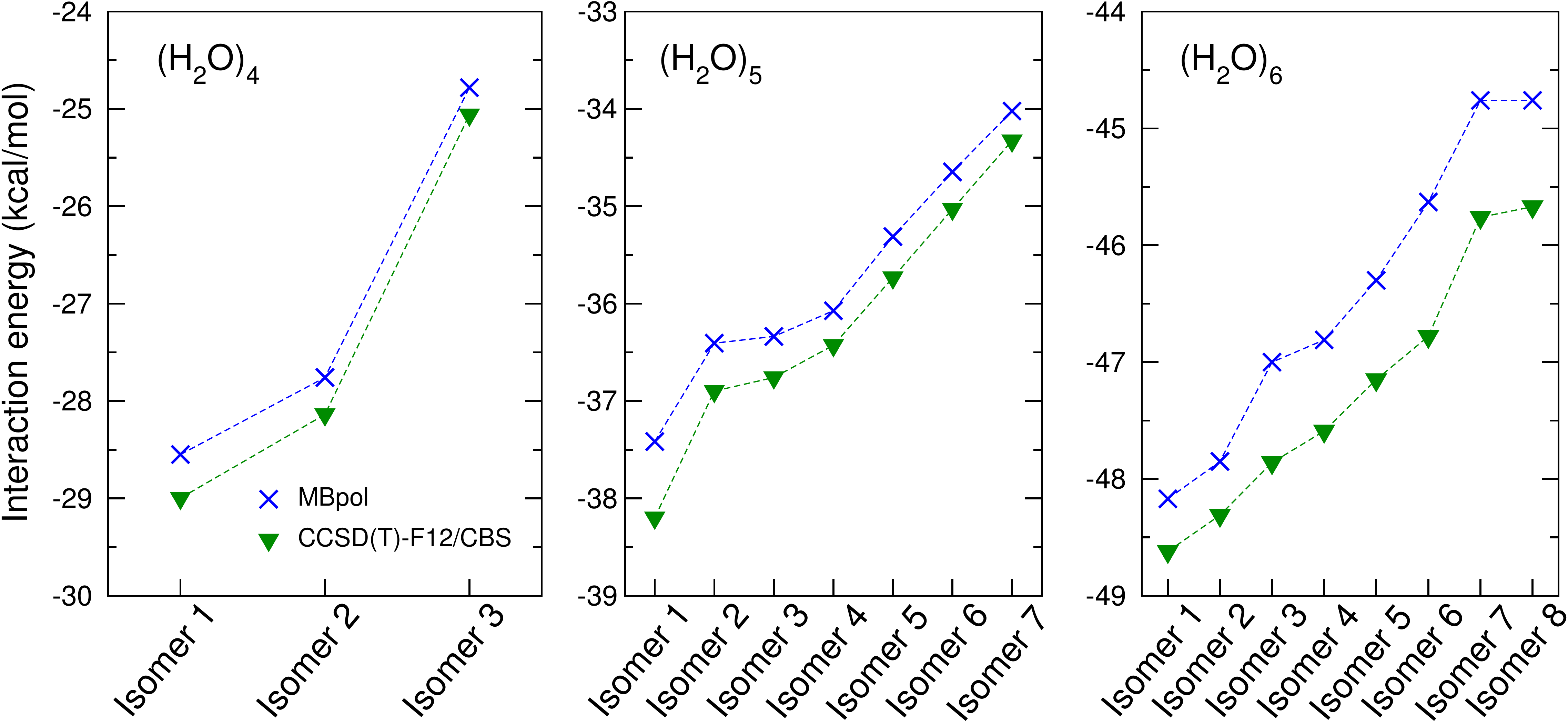}
\end{center}
\caption{Interaction energies of all the low-lying isomers of the 
(H$_2$O)$_\text{n}$ (n=4,5,6) clusters obtained with the MB-pol potential
and the CCSD(T)/CCSD(T)-F12 method in the CBS limit.
}
\label{fig:clusterinterE}
\end{figure}

All (H$_2$O)$_\text{n}$ clusters, with n = 4$-$6, considered in the analysis of many-body
contribution to the interaction energies are shown in Figure~\ref{f:water-clusters}, while
the errors associated with the individual terms are shown in Figure~\ref{fig:clustermbdecomposition}. 
As explained in Section~\ref{section:methods}, the reference data
were obtained with the CCSD(T)/CCSD(T)-F12b method in the CBS limit.
Independently of the cluster size and geometry, MB-pol exhibits small errors, which are always 
within 0.5 kcal mol\inv~relative to the reference values, 
for all terms of the MBE of Eq.~\ref{eq:mbexpansion}. The error in the
total interaction energy increases with system size as the individual errors start to add up, most
prevalently in ring-type configurations that consist of repeating
dimer and trimer units. Due to extended hydrogen bonding
and symmetry, the ring-type isomers also show
larger higher-body contributions that can be non-negligible.\cite{FrancescoJCP2015} 
The comparison between the reference and MB-pol interaction energies
for the tetramer, pentamer, and hexamer isomers shown in Figure~\ref{fig:clusterinterE},
indicates that the relative interaction energy order
for the different isomers of each cluster is retained by MB-pol, with
a maximum deviation in the computed interaction energies of 0.84 kcal mol\inv.
In this analysis, the interaction energy is defined as the cluster energy minus the energy 
of the individual water molecules kept at the same geometry as in the cluster. 
The comparison in Figure~\ref{fig:clusterinterE} thus directly probes the actual interaction 
between water molecules without being affected by differences in the CCSD(T)/CCSD(T)-F12b 
and MB-pol representations of the monomer distortion energies. 

{\it Scoring.}
To quantify the accuracy of MB-pol in
describing many-body interactions, scores are
assigned to each (H$_2$O)$_\text{n}$ cluster based on the following three
quantities. First, the maximum of the total unsigned error of the individual
terms of the MBE, 
$\Delta\text{E}^{\text{MB}}_{\text{unsigned, max}}$, across all isomers
of an $n$-mer, 
\begin{equation}
\Delta \text{E}_{\text{unsigned, max}}^{\text{MB}} =
\max_i\Big[\Big(\sum_{j}^{\text{NB}}\Big|\text{E}^{j}_{\text{MB-pol}} -
  \text{E}^{j}_{\text{Ref.}}\Big|\Big)_{\text{Isomer i}}\Big].
\end{equation}
Second, the maximum absolute error in the total interaction energy, $\Delta\text{E}_{\text{Int, max}}$, 
across all isomers of an $n$-mer,
\begin{equation}
\Delta \text{E}_{\text{Int, max}} =  \max_i\Big[\Big|\Big(\sum_{j}^{\text{NB}}\text{E}^{j}_{\text{MB-pol}} 
- \text{E}^{j}_{\text{Ref.}}\Big)_{\text{Isomer i}}\Big|\Big].
\end{equation}
Third, the relative energy difference between the interaction energy of the
prism and cage hexamer isomers,
\begin{equation}
\Delta \text{E}^{\text{Int}}_{\text{rel}} = \text{E}^{\text{Int}}_{\text{prism}} - \text{E}^{\text{Int}}_{\text{cage}}.
\end{equation}
A small value for $\Delta\text{E}_{\text{Int, max}}$ with a larger error in
$\Delta\text{E}^{\text{MB}}_{\text{unsigned, max}}$ would indicate
that the MBE benefits from error compensation.

For the first two criteria, 100 points are assigned if the magnitude of the error
is less than 1 kcal mol\inv, which is commonly defined as
chemical accuracy. 10 points are deducted successively for every 0.5 kcal 
mol\inv~increase in the error. For the third criterion, the reference value 
calculated using the CCSD(T)/CCSD(T)-F12/CBS method is
$-$0.29 kcal mol\inv. A score of zero is 
assigned for the wrong sign while a score of 100 is assigned if the relative energy
is within 0.1 kcal mol\inv~from the reference value. 10 points are deducted for
every additional 0.1 kcal mol\inv~difference.  
The MB-pol values for the three criteria and associated scores are reported in 
Table~\ref{tab:mberrors}, which demonstrates that MB-pol receives perfect scores 
in all three criteria for (H$_2$O)$_\text{n}$ clusters with n = 4$-$6.

\begin{table}[!htbp]
\caption{Scores to assess the accuracy of MB-pol in describing
many-body interaction energies in small water clusters. Energies are in kcal mol$^{-1}$.
$\Delta \text{{E}}^{\text{MB}}_{\text{unsigned, max}}$ is the maximum
 total unsigned error, $\Delta {\text{E}}_{\text{Int, max}}$ is the
 maximum absolute total error in the
many-body decomposition, $\Delta \text{{E}}^{\text{Int}}_{\text{rel}}$ is the difference
between the interaction energy of the prism and cage hexamer isomers.}
\label{tab:mberrors}

\begin{tabular}{ c c c c c c c } 
\hline
\hline
    Cluster \hspace{0.1cm} & $\Delta\text{E}^{\text{MB}}_{\text{unsigned, max}}$ \hspace{0.1cm} & Score \hspace{0.1cm}   & $\Delta\text{E}_{\text{Int, max}}$ \hspace{0.1cm} & Score \hspace{0.1cm} &  $\Delta\text{E}^{\text{Int}}_{\text{rel}}$ \hspace{0.1cm} & Score \hspace{0.1cm} \\ 
\hline
    (H$_2$O)$_4$ & 0.26 & 100  & 0.26 & 100 & -         & -  \\ 
    (H$_2$O)$_5$ & 0.72 & 100  & 0.72 & 100 & -         & -    \\ 
    (H$_2$O)$_6$ & 0.84 &  100  & 0.84 &  100  & -0.32  & 100   \\ 

\hline
\hline
\vspace{0.1cm}
\end{tabular}
\end{table}

\subsubsection{Vibrational frequencies}
\label{subsection:vib_freq}
The ability of MB-pol in predicting accurate vibrational frequencies
is assessed through the analysis of harmonic frequencies calculated for
small water clusters, from the water monomer to the hexamer.
The comparisons are made with recently published benchmark
data that are expected to closely approximate 
CCSD(T)/CBS values.\cite{Howard2014,Howard2015}

The water clusters included in this analysis are 
all structures with lowest energy for n = 2 - 6, along with the cage, book1, 
and cyclic-chair (also referred to as ring) isomers of the hexamer
(see Figure \ref{f:water-clusters}).
For n = 3 - 5, the minimum energy structures correspond
to cyclic structures in which each water molecule donates and
accepts one hydrogen bond.
In the water hexamer the lowest energy structure is the prism isomer,
which is nearly isoenergetic with the cage isomer (see Section~\ref{subsection:mbe}), 
while the book1 and cyclic-chair isomers are higher in energy. Prism and cage are the
predominant isomers at very low temperatures while the cage and cyclic-chair isomers
appear at higher temperatures.
\cite{nauta_k_2000, burnham_cj_2002, elliot_bm_2008, saykally_rj_2012, 
perez_c_2012, wang_y_2012, babin_v_2013} 

\renewcommand{\arraystretch}{1.4}
\begin{table}[t]
  \caption{Average deviation (AD), average absolute deviation (AAD),
    and maximum absolute deviation (MAD) of MB-pol harmonic
    frequencies (in cm$^{-1}$) from benchmark data for the water
    monomer and (H$_2$O)$_\textrm{n}$ clusters (n = 2$-$6). Scores are assigned
    based on the MAD value (see text).}
  \label{t:freqs}
  \begin{ruledtabular}
  \begin{tabular}{lccccccccc}
    & \mcone{H$_2$O} & \mcone{(H$_2$O)$_2$} & \mcone{(H$_2$O)$_3$}
    & \mcone{(H$_2$O)$_4$} & \mcone{(H$_2$O)$_5$}
    & \mcone{cyclic-chair} & \mcone{book1} & \mcone{cage} & \mcone{prism} \\
    \hline
    AD  & -1.3 & -1.8 & -1.7 & -5.1 & -6.0 & -7.3 & -5.4 & -3.3 & -2.8 \\
    AAD &  1.3 &  4.4 &  4.7 & 11.6 & 15.6 & 16.5 & 12.7 &  8.9 &  7.8 \\
    MAD &  2.2 & 12.0 & 16.5 & 31.4 & 46.2 & 49.4 & 38.6 & 23.0 & 24.1 \\
    Score &  100 & 100 & 100 & 90 & 80 & 80 & 90 & 90 & 90    \\         
  \end{tabular}
  \end{ruledtabular}
\end{table}

Table \ref{t:freqs} reports the average deviation (AD), average absolute
deviation (AAD), and the maximum absolute deviation (MAD) between the reference
and the MB-pol harmonic frequencies calculated for the water monomer and each of
the eight water clusters.  The complete list of harmonic frequencies can be
found in the Supplementary Material.  In all cases, MB-pol accurately
reproduces the reference data, with AAD and MAD values never exceeding
17~cm$^{-1}$ and 50~cm$^{-1}$, respectively.  The MB-pol deviations from the
reference data increase with increasing cluster size, in particular for cyclic
structures, in which, as discussed in Section~\ref{subsection:mbe}, 
small errors in the MBE can add up for repeating
units due to the inherent symmetry.  As a result, the deviations for the
hexamer book1, cage, and prism isomers are smaller than for the cyclic-chair
isomer. On average, the MB-pol harmonic frequencies are slightly redshifted as
compared to the reference data.  This redshift is mostly due to the
low-frequency intermonomer and bending modes, while the OH stretch frequencies
are slightly blueshifted (see list of frequencies in the Supporting
Information).  Both red and blue shifts associated, respectively, with the bending and stretching
vibrations can be explained by considering that MB-pol does not allow
water autoionization, and, consequently, underestimates the ionic character, and thus
the strength, of the hydrogen bonds. 
Importantly, it has been shown that MB-pol provides a consistent representation 
of the vibrational structure of water independently of the system size, correctly
predicting infrared,\cite{FrancescoJCTC2015} 
Raman,\cite{FrancescoJCTC2015} sum-frequency generation,\cite{FrancescoJACS2016} and 
two-dimensional infrared\cite{FrancescoJPCB2016} spectra of liquid water
at ambient conditions.

{\it Scoring.}
To quantify the accuracy with which MB-pol reproduces the reference harmonic frequencies,
a score for each cluster is assigned based on the corresponding MAD value. 
A score of 100 is assigned if the MAD is below the threshold value of
20 cm\inv, and 10 points are then deducted for each additional increment of 20 cm\inv.
The MB-pol scores are reported in Table~\ref{t:freqs}.

\subsection{Thermodynamic properties of liquid water}
The accuracy of MB-pol in reproducing the properties of liquid 
water is assessed through comparisons with the corresponding experimental data
as a function of temperature. As discussed in detail in Section~\ref{section:methods},
all MB-pol properties were calculated from classical MD simulations. 
The role played by NQE (e.g., zero-point energy and tunneling)
will be discussed in a forthcoming publication.  
A summary of several thermodynamic properties of liquid water computed with MB-pol
at atmospheric pressure (P = 1 atm) along
with the corresponding experimental values is reported in the Supplementary Material
(Tables XL and XLI).

\subsubsection{Density} 
\label{subsection:density}
\begin{figure}
\begin{center}
\includegraphics[width=0.75\textwidth]{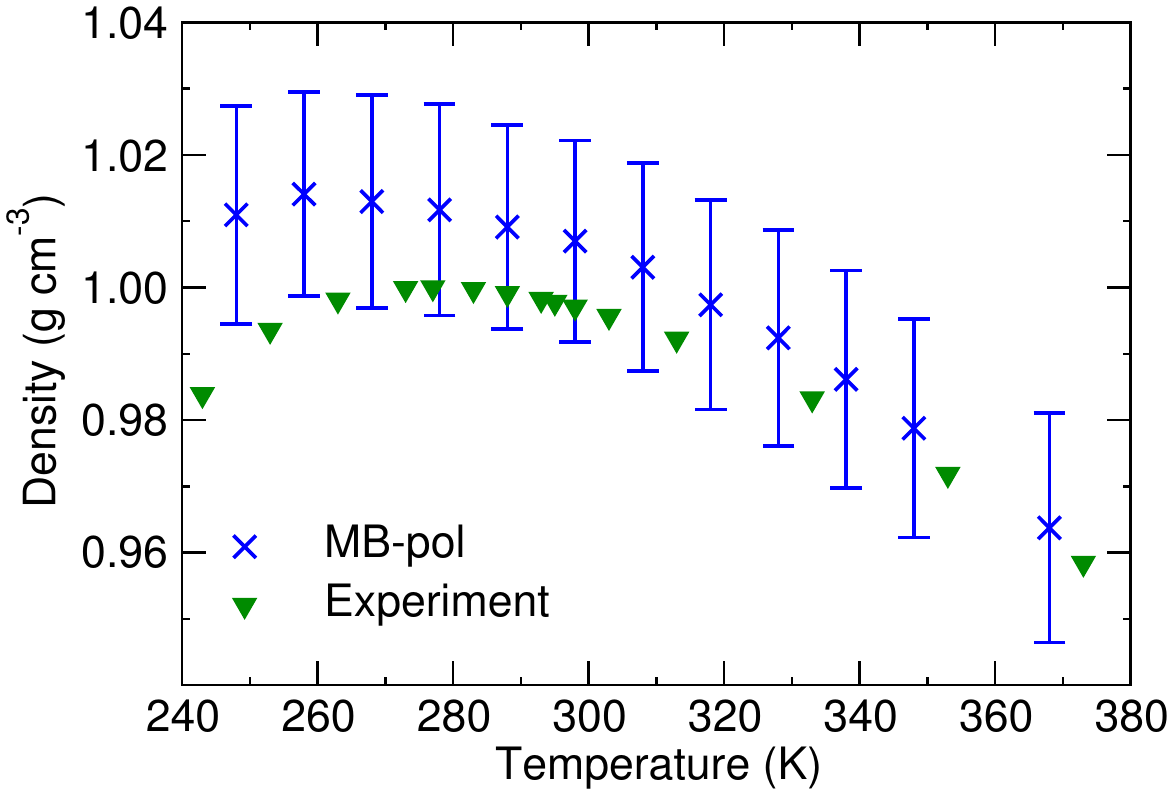}
\end{center}
\caption{Densities of liquid water at atmospheric pressure from MB-pol. Experimental
	data are taken from Refs.\citenum{Speedy1987} and \citenum{W.WagnerandA.Prub2002}.
}
\label{fig:ldensity}
\end{figure}

The temperature dependence of the density of
liquid water at 1 atm calculated from classical MD simulations with MB-pol is shown in
Figure~\ref{fig:ldensity}. The experimental data are taken from
Refs.~\citenum{Speedy1987} and \citenum{W.WagnerandA.Prub2002}. 
At high temperature, the MB-pol results 
are in excellent agreement with the corresponding experimental values.  
As the temperature decreases, the difference between MB-pol and experiment
increases nearly linearly. 
The maximum and average absolute deviations from the reference 
values are 0.042 and 0.013 g cm$^{-3}$, respectively.  
The temperature of maximum density, obtained by calculating the analytical
derivative of a fifth-order polynomial interpolating
the simulation results, is located at 263 K, which is
14 K lower than experiment.\cite{Speedy1987,W.WagnerandA.Prub2002}
The systematic deviation between the MB-pol and the experimental values 
as the temperature decreases can be attributed, at least in part, to the neglect
of NQE in the simulations, which, as expected, become
increasingly important at lower temperature.\cite{CeriottiChemRev2016}
In this context, at 298~K, the density predicted by classical MD simulations with MB-pol
is 1.007 g cm$^{-3}$ compared to the experimental value of 0.997 g cm$^{-3}$. 
Using path-integral molecular dynamics (PIMD) simulations, which
explicitly include NQE, it was shown that
the density of water predicted by MB-pol at the same temperature decreases 
by 0.6\%,\cite{Francesco1JCTC2014} in excellent agreement with the experimental value.

\subsubsection{Enthalpy of vaporization }
The enthalpy of vaporization, $\Delta H_{vap}$, is one of the properties usually 
included in the fitting procedures to parameterize empirical water potentials
\cite{tip4p_Vega2005,tip4p_Gordon2004} and is defined as
\begin{equation}
	\Delta H_{vap} = H_{gas}-H_{liquid} = U_{gas}-U_{liquid}+P(V_{gas}-V_{liquid})
	\label{eq:enthalpy},
\end{equation}
where $H$, $U$, and $V$ are the enthalpy, internal energy, and volume, respectively,
and the subscripts denote that the molecules are in either the gas or liquid state. 
Since, at low pressure, the gas can be considered ideal, the contribution to $U_{gas}$
due to interactions between water molecules can be neglected and $\Delta H_{vap}$
can then be rewritten (for 1 mol of water) as
\begin{equation}
	\Delta H_{vap} = U_{gas}-U_{liquid}-PV_{liquid}+RT ,
\end{equation}
where $U_{gas}$ contains the average kinetic (i.e., 3/2 RT) and potential (i.e., intramolecular
distortion) energies of the gas molecules. At each temperature, the average potential energy was 
calculated from a 2 ns long classical MD simulation of a single water molecule.
It should be noted that, unlike point charge models, $\Delta H_{vap}$ calculated with MB-pol
implicitly includes the self-polarization correction,\cite{tip4p_Vega2005,tip4p_Gordon2004}
since MB-pol correctly describes the change in dipole moment from the gas to the condensed phase.

\begin{figure}
\begin{center}
\includegraphics[width=0.75\textwidth]{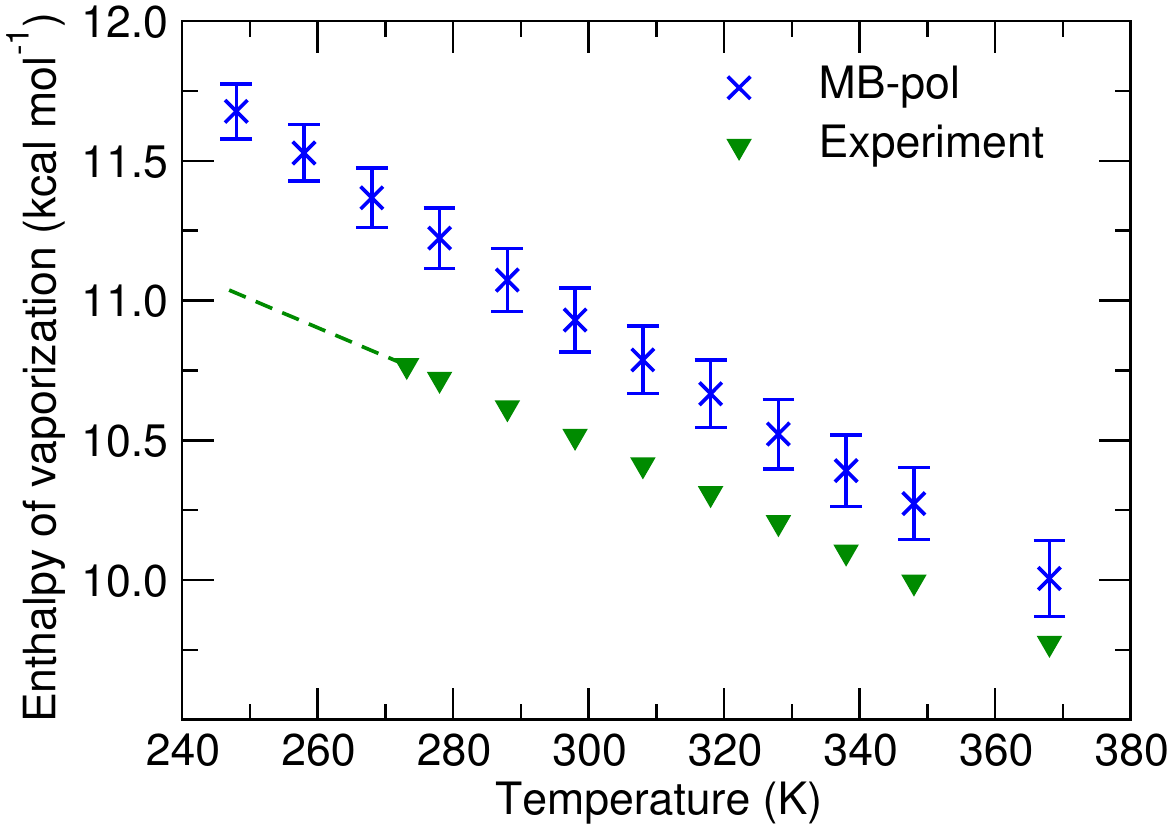}
\end{center}
\caption{Enthalpy of vaporization of liquid water at atmospheric pressure calculated
using MB-pol. Experimental data is provided for
comparison.\cite{W.WagnerandA.Prub2002} The dashed line corresponds to the
extrapolated data. 
}
\label{fig:lenthalpy}
\end{figure}

Figure~\ref{fig:lenthalpy} shows that $\Delta H_{vap}$ calculated from classical MD simulations
with MB-pol is systematically larger than the corresponding experimental values, with
an average absolute deviation of 0.41 kcal mol\inv.
As for the density, the deviation from experiment increases as the temperature decreases, which
can be attributed to the neglect of NQE in the simulations.  
At 298~K, $\Delta H_{vap}$ predicted by MB-pol is 10.93 kcal mol\inv, 0.42 kcal mol\inv~larger 
than the experimental value. Guillot and Guissani suggested that $\Delta H_{vap}$ 
for hypothetical ``classical water" at 298.15 K should be 11.0 kcal mol\inv,\cite{GuillotJCP2001}
which is in excellent agreement with the value obtained from classical MD simulations with MB-pol.

\begin{figure}[t]
\begin{center}
\includegraphics[width=0.75\textwidth]{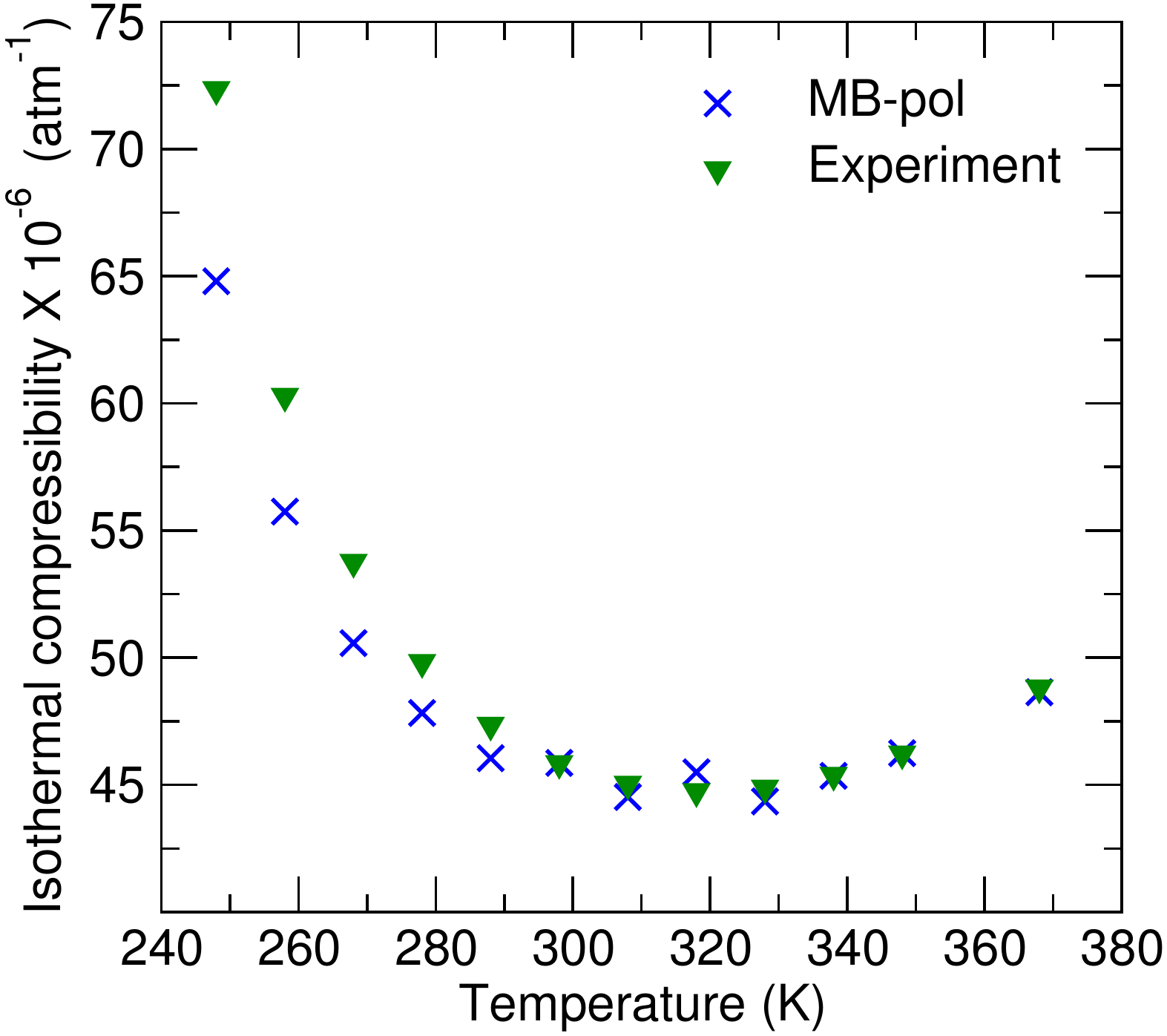}
\end{center}
\caption{Isothermal compressibility of liquid water at atmospheric pressure
calculated using MB-pol. Experimental data
are taken from Ref.\citenum{Kell1975}.
}
\label{fig:lcompressibility}
\end{figure}

\subsubsection{Isothermal compressibility} 
The isothermal compressibility, $\kappa_T$, is defined as
\begin{equation}
	\kappa_T =  -{\frac{1} {V}} \left({\frac{\partial V}{\partial P}}\right)_T,
	\label{eq:lcompressibility}
\end{equation}
where $V$ and $P$ are the volume and the pressure of liquid water at a given
temperature $T$, respectively. 
To solve Eq.~\ref{eq:lcompressibility}, classical MD simulations 
with MB-pol were carried out at each temperature for seven pressure values 
(-1.5, -1.0, -0.5, 0.001, 0.5, 1.0, 1.5 katm), which were used to determine
the corresponding average volumes. 
$\kappa_T$ was then calculated from the derivative of a third-order polynomial
that was used to fit the volumes as a function of pressure at each temperature.
As shown in Figure~\ref{fig:lcompressibility}, the MB-pol values are in good 
agreement with the corresponding experimental data,\cite{Kell1975}
correctly predicting a minimum between 310 and 330 K.
The average and maximum absolute deviations from experiment in the temperature 
range between the freezing and the boiling point are 
0.5$\times$10$^{-6}$ and 1.8$\times$10$^{-6}$ atm\inv,
respectively. As expected, the deviations from the experimental
data become more pronounced
at low temperature due to the neglect of NQE in classical MD simulations.

\begin{figure}[h]
\begin{center}
\includegraphics[width=0.75\textwidth]{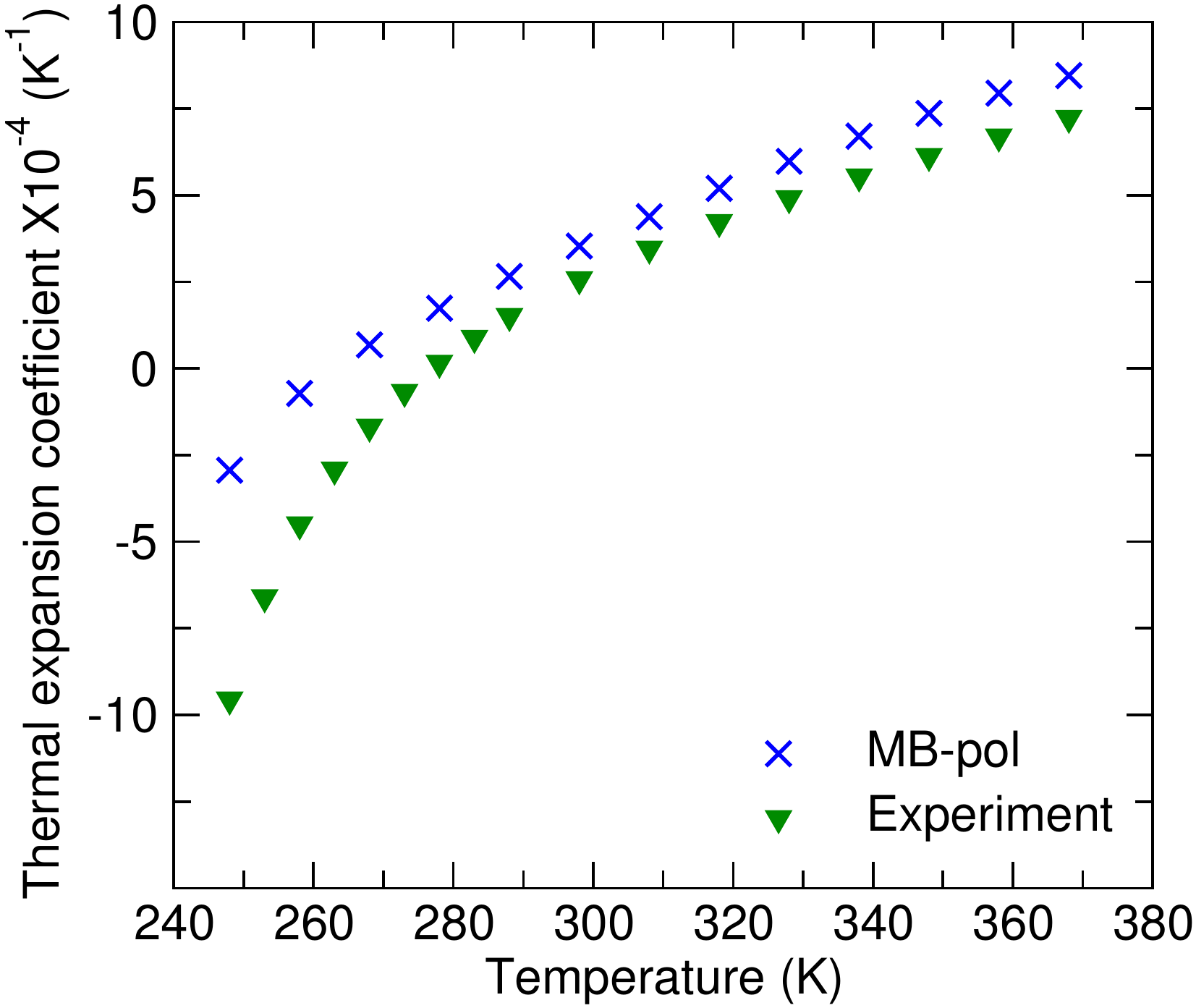}
\end{center}
\caption{Thermal expansion coefficient of liquid water at atmospheric pressure. Experimental data
are taken from Ref.\citenum{Kell1975}.
}
\label{fig:lthermal}
\end{figure}

\subsubsection{Thermal expansion coefficient}
The thermal expansion coefficient, $\alpha_P$, is defined as
\begin{equation}
	\alpha_P =  -{\frac{1}{V}} \left({\frac{\partial V}{\partial T}}\right)_P 
	= -{\frac{1}{\rho}} \left({\frac{\partial \rho}{\partial T}}\right)_P
	\label{eq:expansion_coeff1}.
\end{equation}
where $V$, $T$, and $\rho$ are the volume, the temperature, and
the density of liquid water at a given pressure $P$, respectively.  From
Eq.~\ref{eq:expansion_coeff1}, it is evident that any water model that
accurately predicts the temperature dependence of the density is also capable
of correctly reproducing the variation of $\alpha_P$.  The MB-pol values of
$\alpha_P$, shown in Figure~\ref{fig:lthermal} along with the corresponding
experimental data,\cite{Kell1975} were determined numerically by solving
Eq.~\ref{eq:expansion_coeff1} using the density values reported in
Figure~\ref{fig:ldensity}.  Similar to $\rho$, $\alpha_P$ derived from
classical MD simulations with MB-pol is in close agreement with experiment
above the freezing point, with average and maximum absolute deviations of
1.3$\times$10$^{-4}$ and 1.7$\times$10$^{-4}$ K\inv, respectively.  At the
classical level, MB-pol predicts $\alpha_P$ to be zero at 263 K, in line with
the density maximum reported in Section~\ref{subsection:density}
(experimentally, $\alpha_P = 0$ at 277~K).  Again, NQE
appear to be important to quantitatively reproduce the variation $\alpha_P$ at
low temperature, as already found in previous sections for other thermodynamic
properties of liquid water.

\subsubsection{Isobaric heat capacity}

\begin{figure}
\begin{center}
\includegraphics[width=0.75\textwidth]{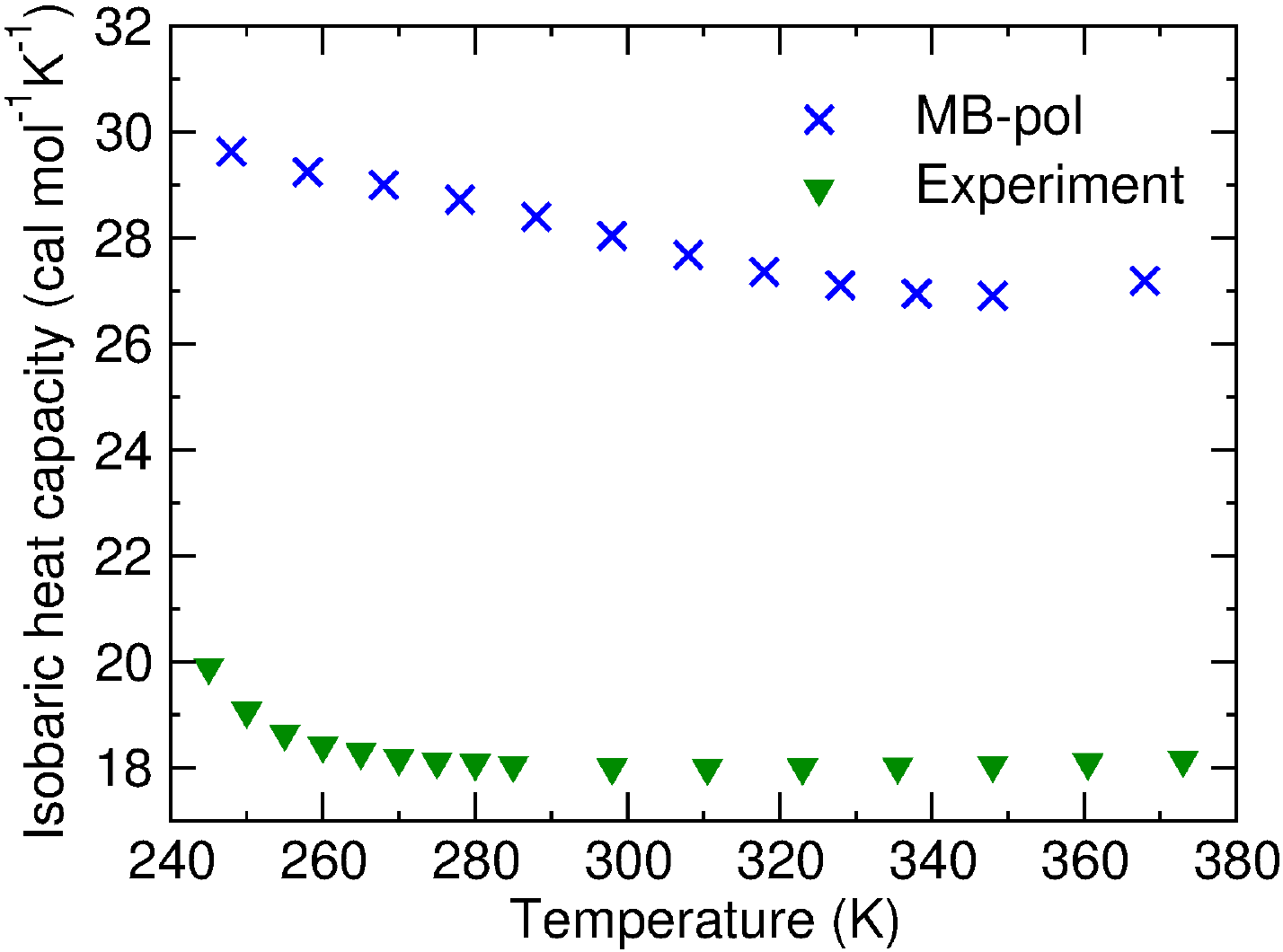}
\end{center}
\caption{Isobaric heat capacity of liquid water at atmospheric pressure. 
	Experimental data are taken from Refs.\citenum{W.WagnerandA.Prub2002} and \citenum{Archer2000}.
}
\label{fig:lheat}
\end{figure}

The isobaric heat capacity, c$_P$, is defined as
\begin{equation}
	c_P = \left({\frac{\partial H}{\partial T}} \right)_P,
	\label{eq:lheat1}
\end{equation}
where $H$ and $T$ are the enthalpy and the temperature of liquid water
at a given pressure $P$, respectively. 
It should be noted that, unlike other thermodynamic properties,
$c_P$ is significantly affected by NQE even at room
temperature.\cite{ShinodaJCP2005,VegaJCP2010}
Using Eq.~\ref{eq:lheat1}, $c_P$ was calculated as the
temperature derivative of a fifth-order polynomial interpolating the values of
$H$ obtained at different temperatures from classical MD simulations carried out
in the NPT ensemble.  
Consistent with previous studies, the classical MB-pol results shown 
in Figure~\ref{fig:lheat}  are larger than the corresponding experimental data
\cite{W.WagnerandA.Prub2002,Archer2000} at all temperatures.
Experimentally, it is known that the
difference between the heat capacity of liquid H$_2$O and D$_2$O increases as
the temperature decreases.\cite{Angell1982} For example, the heat capacity of
H$_2$O and D$_2$O at 250 K are 19.07 and 23.93 cal mol\inv~K\inv,
respectively.\cite{W.WagnerandA.Prub2002, Archer2000,Angell1982} The
corresponding value obtained from classical MD simulations with MB-pol is 29.25
cal mol\inv~K\inv, reinforcing the notion that explicit inclusion of NQE
in simulations with MB-pol is required for quantitative, and
physically correct, calculations of $c_P$ at all temperatures. Levitt et al.
estimated that 6 cal mol\inv~K\inv~ must be subtracted from the classical value 
of the heat capacity at constant volume, $c_V$, to account for NQE at 298.15~K.
\cite{LevittJPCB1997} Assuming that the same correction can be applied to $c_P$, 
the quantum-corrected MB-pol result at 298~K becomes 21.85 cal mol\inv~K\inv, in closer 
agreement with the experimental value.

\subsubsection{Surface tension}
\begin{figure}
\begin{center}
	\includegraphics[width=0.75\textwidth]{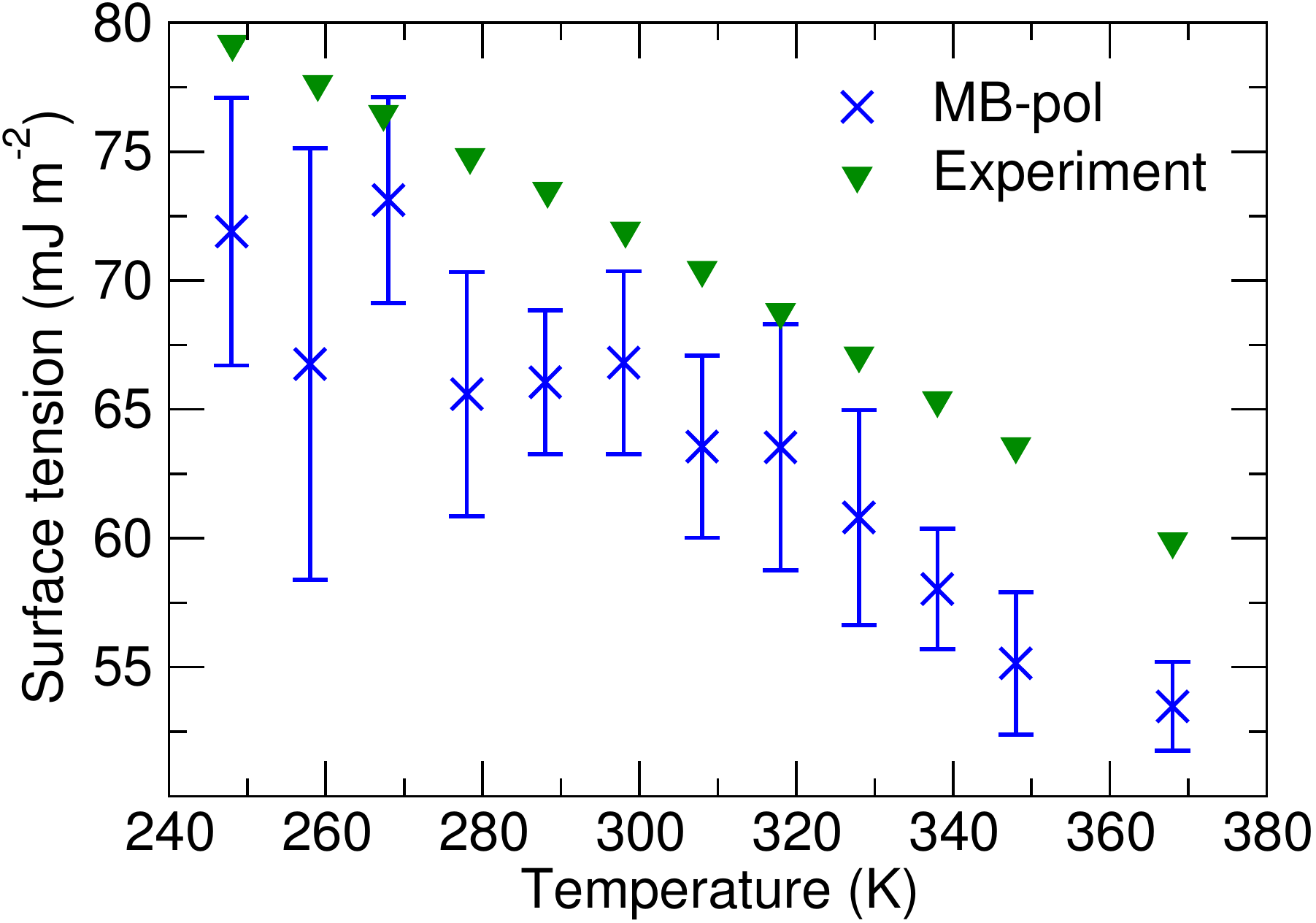}
\end{center}
\caption{Surface tension of liquid water at atmospheric pressure. Experimental data
	are taken from Refs.\citenum{VoljakJPCRD1983} and \citenum{JanJPCB2015}.
}
\label{fig:lsurface}
\end{figure}

The surface tension, $\gamma$, can be calculated from MD simulations as\cite{Widom1982}
\begin{equation}
\gamma = \int_{z_\alpha}^{z_\beta} \left[P_N(z)-P_T(z)\right]dz,
\label{eq:surface_tension1}
\end{equation}
where the z-direction is defined along the normal vector to the surface, $\alpha$ and $\beta$ 
refer to the liquid and the gas phase, respectively, and N and T denote the normal and tangential
components of the pressure tensor, respectively. 
Considering the slab geometry defined in Section~\ref{section:methods}, Eq.~\ref{eq:surface_tension1}
reduces to
\begin{equation}
\gamma = \frac{1}{2} L_z \left<P_{zz} - \frac{1}{2}\left(P_{xx}+P_{yy}\right)\right> ,
\label{eq:surface_tension2}
\end{equation}
where $P_{xx}$, $P_{yy}$, and $P_{zz}$ are the diagonal elements of the pressure tensor, 
$L_z$ is the length of the simulation box in the z-direction, 
and the angular brackets denote an ensemble average.
The comparison between the experimental\cite{VoljakJPCRD1983,JanJPCB2015} 
and MB-pol values of the surface tension is shown in Figure~\ref{fig:lsurface}.
Overall, the MB-pol results are in good agreement with experiment over the entire 
temperature range considered in this study, correctly reproducing (within statistical uncertainty)
the linear increase of $\gamma$ as the temperature decreases.
At 298 K, the surface tension obtained from classical MD simulations with MB-pol is 
$66.8\pm3.55$ mJ m$^{-2}$ compared to the experimental value of
71.97 mJ m$^{-2}$.\cite{VoljakJPCRD1983}
Contrary to other thermodynamic properties, the differences between the experimental and MB-pol
values of the surface tension remain effectively constant as a function of temperature, suggesting
that $\gamma$ is less sensitive to NQE.

\subsubsection{Static dielectric constant}
\begin{figure}
  \begin{center}
    \includegraphics[width=0.75\textwidth]{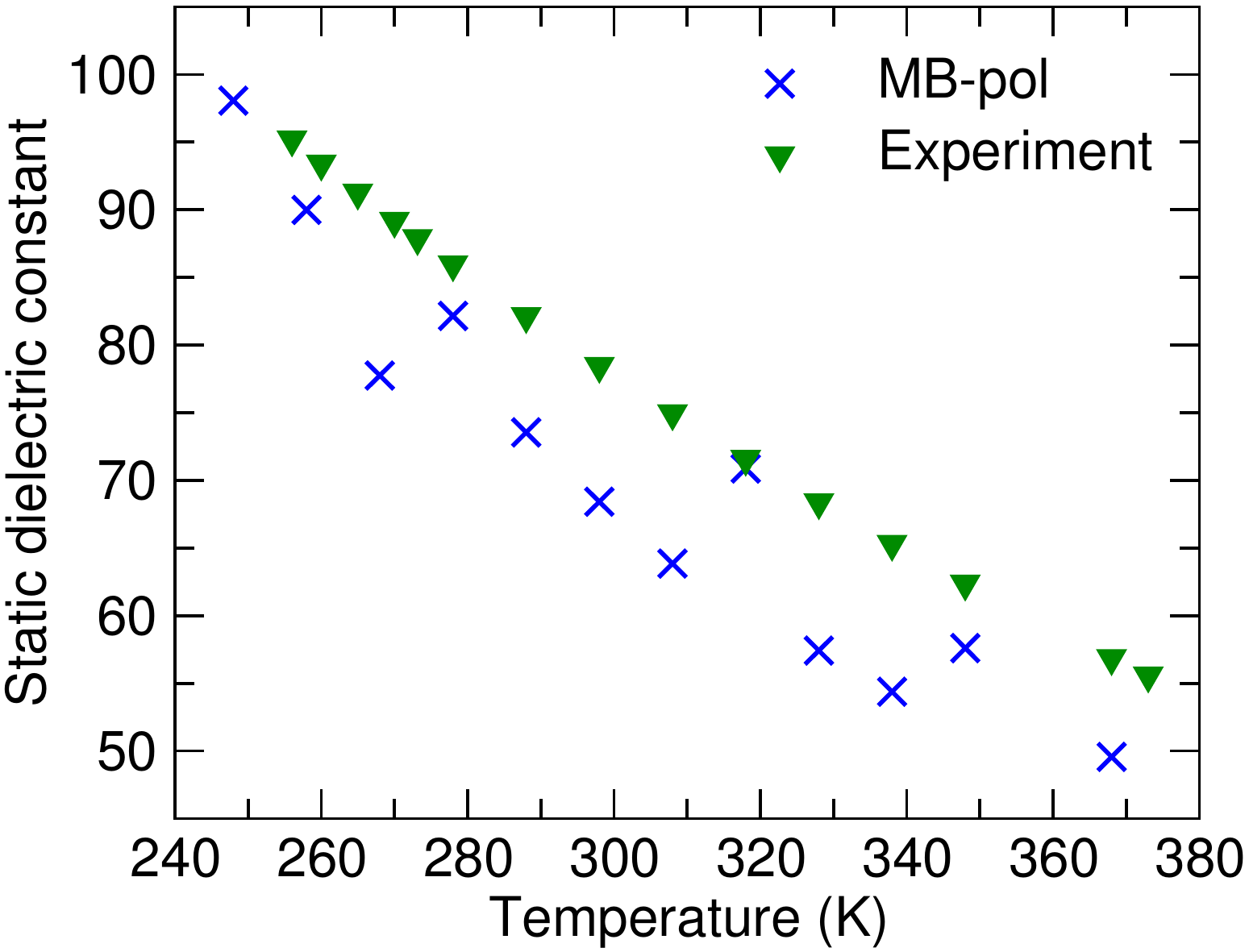}
  \end{center}
  \caption{Static dielectric constant of liquid water at atmospheric pressure. Experimental data
    are taken from Ref.\citenum{Williams1997}.
  }
  \label{fig:ldielectric}
\end{figure}

The static dielectric constant, $\epsilon$, is defined as
\begin{equation}
  \epsilon = 1 + \frac{4\pi}{3Vk_B\left<T\right>}\left(\left<\mathbf M^2\right>-\left<\mathbf M\right>^2\right),
  \label{eq:dielectric}
\end{equation}
where $\textbf{M}$ is the total dipole moment of the simulation box, $k_B$ is
Boltzmann's constant, $V$ and $T$ are the volume and the temperature, respectively,
and the angular brackets denote an ensemble average.
The temperature dependence of the dielectric constant
calculated from classical MD simulations with MB-pol
is compared in Figure~\ref{fig:ldielectric} with the corresponding experimental
data.\cite{Williams1997}
$\epsilon$ calculated from the MB-pol simulations is in good agreement with
experiment over the entire temperature range considered in this study.
At 298~K, the value of $\epsilon$ obtained from the MB-pol simulations is 68.4
which is $\sim$13\% smaller than the experimental value of 78.5. These differences
may be related, at least in part, to (small) differences in the liquid structure due to the
neglect of NQE,\cite{Francesco1JCTC2014} which, in turn, may affect the dipole moments
of the water molecules.

\subsubsection{Self-diffusion coefficient} 
\begin{figure}[h]
\begin{center}
\includegraphics[width=0.75\textwidth]{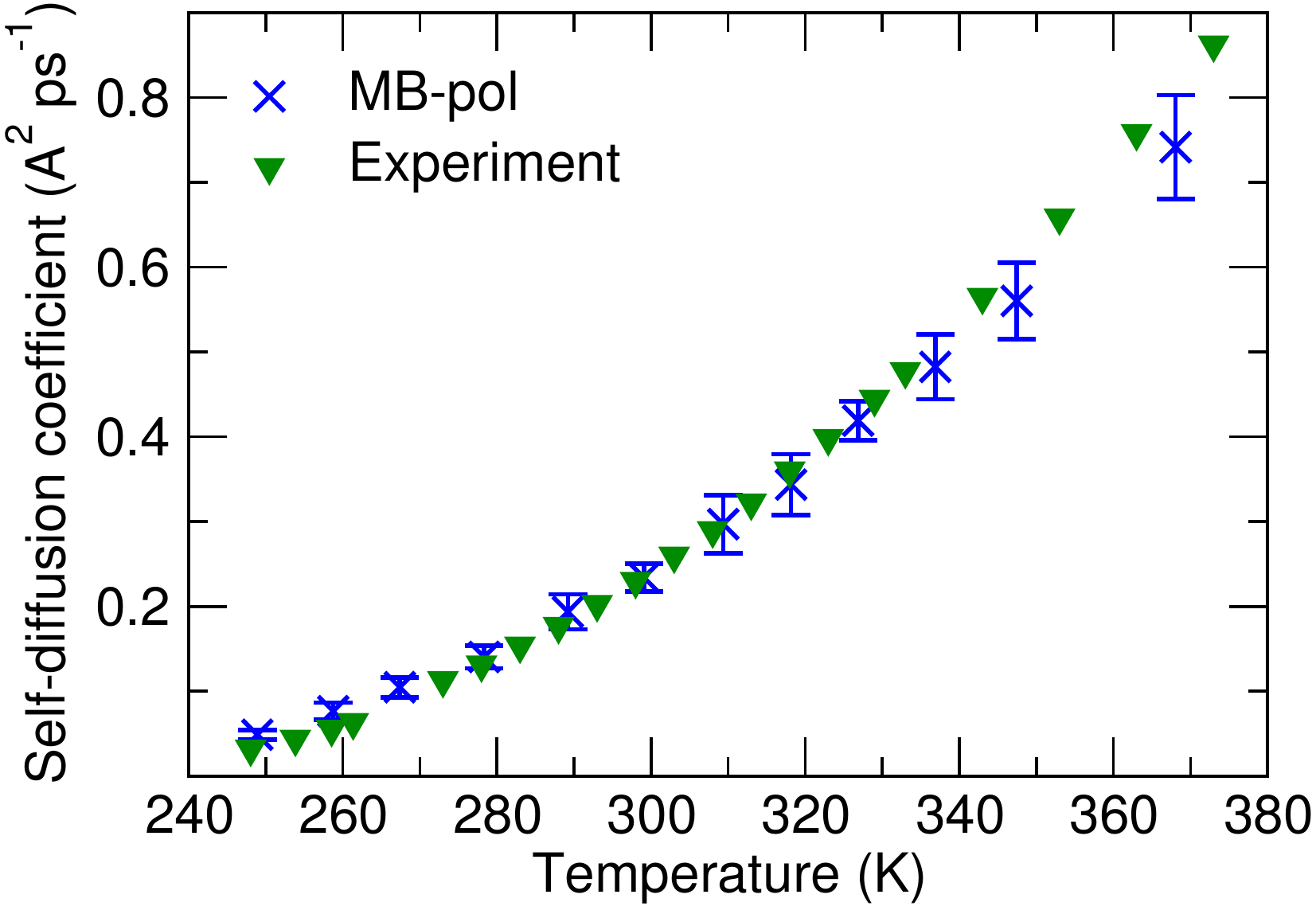}
\end{center}
\caption{Self-diffusion coefficient of liquid water at atmospheric pressure. Experimental data
are taken from Refs.\citenum{HochJCP1972,SaccoPCCP2000,WoolfJCSFT1989,MillsJPC1973}.
}
\label{fig:ldiffusion}
\end{figure}

The self-diffusion coefficient, D, of liquid water can be computed
from the velocity autocorrelation function as
\begin{equation} 
D = \frac{1}{3} \int_{0}^{\infty} \left<v_i(t)v_i(0)\right> dt ,
\end{equation} 
where $v_i$ is the velocity of the center of mass of $i^{th}$ water molecule and
the brackets indicate an ensemble average in the microcanonical (NVE) ensemble.
The MB-pol results are compared in
Figure~\ref{fig:ldiffusion} with the corresponding experimental values.
\cite{HochJCP1972,SaccoPCCP2000,WoolfJCSFT1989,MillsJPC1973} At 298~K,
classical MD simulations with MB-pol predict $D$ to be $0.23\pm0.02$
\AA$^2$ ps\inv~in agreement with the experimental value of 0.229 \AA$^2$ ps\inv.  It
should be mentioned that an incorrect value for the classical MB-pol diffusion
coefficient at room temperature was previously reported by some of us in Ref.
\citenum{Francesco1JCTC2014}.
The similarity between the present classical value and the corresponding
quantum value obtained in Ref.~\citenum{Francesco1JCTC2014} from centroid
molecular dynamics simulations suggests the presence of competing NQE 
in the water diffusion as originally suggested in
Ref.~\citenum{ManolopoulosJCP2009}.  

Since it was shown that the calculation of
the self-diffusion coefficient from MD simulations is particularly sensitive to
the system size,\cite{HummerJPCB2004} additional NVE simulations were carried
out for a system containing 512 molecules. For this larger system, 
$D = 0.24\pm0.02$ \AA$^2$ ps\inv~at room temperature. The self-diffusion coefficient in the
limit of an infinite system size can be calculated as\cite{KremerJCP1993,HummerJPCB2004}
\begin{equation}
  D(\infty) = D(L)+\frac{\xi k_BT}{6\pi \eta L} ,
  \label{eq:ldiffusioncorrection} 
\end{equation}
where $k_B$ is Boltzmann's constant, $T$ is the temperature, $\eta$ is the
viscosity, and $\xi$ is a constant that depends on the shape of the
simulation box (for cubic, $\xi=2.837297$).
Plugging the values obtained for 256 and 512 water molecules into Eq.~\ref{eq:ldiffusioncorrection},
and using the experimental value for $\eta$,
$D(\infty) = 0.280\pm0.040$ \AA$^2$ ps\inv~at 298~K.

\begin{table}[t]
\caption{\label{tab:lscoring}
Scores calculated for MB-pol at 248, 298, and 360 K based on threshold
criteria (x$_{tol}$) set for each property. See text for more details. 
Interpolated data are used at 360 K.
$\rho$ is the density; c$_P$ is the heat capacity;
$\Delta H_{vap}$ is the enthalpy of vaporization; $\kappa_T$ is the isothermal
compressibility; 
$\epsilon$ is the dielectric constant; D is the self-diffusion coefficient; $\gamma$ is the
surface tension. 
}
\begin{ruledtabular}
\begin{tabular}{ccccc}
Property & x$_{tol}$ (\%) & \multicolumn{3}{c}{Scores} \\
&  & 248~K & 298~K & 360~K \\ 
\hline
$\rho$           & 0.5 & 60 & 80 & 100   \\
$\Delta H_{vap}$ & 2.5 & 80 & 90 & 100   \\
$\kappa_T$            & 5   & 80 & 100 & 100   \\
$\alpha_P$            & 0.5   & 90 & 100 & 100  \\
c$_P$            & 5   & 90 & 100 & 100  \\
$\gamma$         & 2.5 &  70 &  80  &  60    \\
$\epsilon$       &  5  &  100 & 80 & 80    \\
$D$       &  5  & 0  & 100 & 90    \\
Average         &   &  71 &  91  &  91    \\
\end{tabular}
\end{ruledtabular}            
\end{table}

{\it Scoring.} 
As in Section~\ref{subsection:clusters}, scores are assigned to each property of liquid water
calculated with MB-pol. Following the procedure described in
Refs.~\citenum{VegaPCCP2011},~\citenum{VegaFarad2009}, and
\citenum{AngelosJCP2016}, the relative error ($\delta x$) associated with each
property was used to define the corresponding score as
\begin{equation} 
	\text{score} = \mathrm{max} \left\{ 100-\mathrm{int}(\delta
x/x_{tol}) \times 10 ,0 \right\}, \label{eq:lscore} 
\end{equation}
where $\delta x = |x_{sim}-x_{exp}|/|x_{exp}|$ and $x_{tol}$ is the threshold
value. The score is 100 if the $\delta x$ value is between 0 and $x_{tol}$.
The score is thus reduced by 10\% for each successive increment by x$_{tol}$.
A score of 100 indicates perfect agreement with experiment while a score of 0
indicates poor performance.  The $x_{tol}$ value for each property was taken
from Ref.~\citenum{VegaPCCP2011}, except for $\alpha_P$ that was not included
in that study.  The scores were calculated at three
different temperatures.  The first temperature selected to assess the
performance of MB-pol is 248~K since
the maximum absolute deviations for all properties calculated from
classical MD simulations with MB-pol are observed at this temperature.  As
mentioned above, since NQE are not included in the present
MB-pol simulations, it is not surprising that the computed values deviate
significantly from experiment, especially at low temperatures.  The performance
of MB-pol is also assessed at room temperature (298~K) and at higher
temperature (360~K). Polynomial fits were used to obtain the values 
of all properties at 360 K.
Table~\ref{tab:lscoring} shows the scores obtained from classical MD
simulations with MB-pol for all properties discussed in the previous
sections. The average scores of MB-pol are 71, 91, and 91 
at 248, 298, and 360 K, respectively.

\subsection{Structural properties of liquid water}

\begin{figure}[t]
\begin{center}
\includegraphics[width=0.75\textwidth]{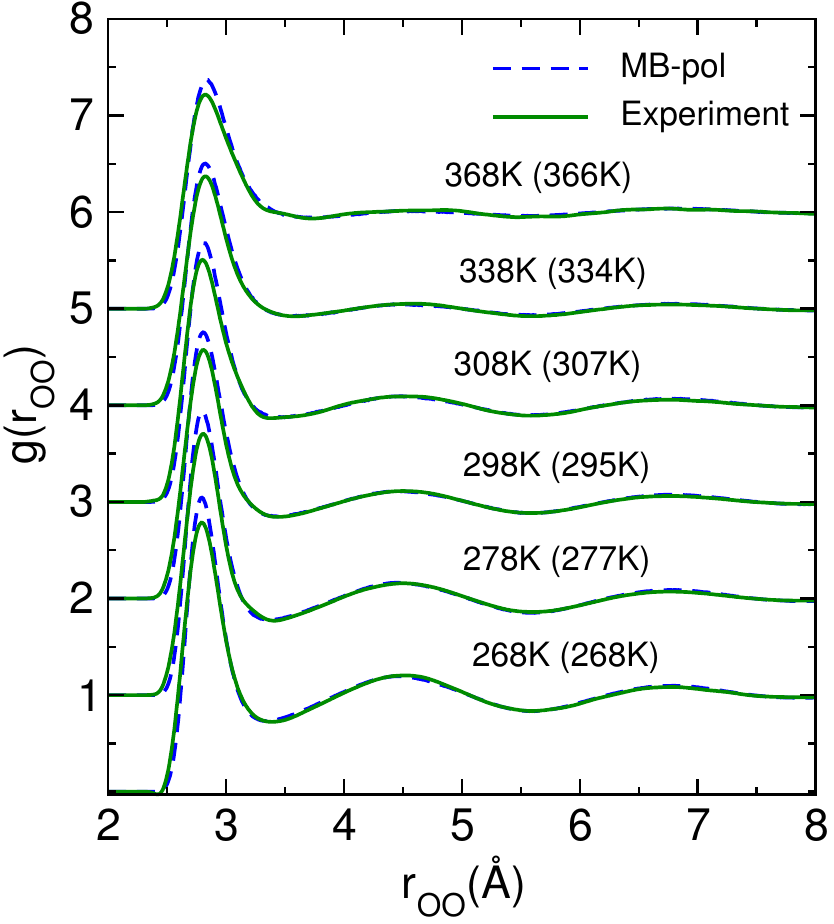}
\end{center}
\caption{Temperature dependence of the oxygen-oxygen RDF
of liquid water predicted by classical NPT simulations with MB-pol 
compared with the corresponding results derived
from X-ray diffraction measurements.\citenum{SkinnerJCP2013,SkinnerJCP2014}.
The temperatures at which the experimental measurements 
were performed are given in parenthesis.
}
\label{fig:lpair_corr}
\end{figure}

\begin{figure}
\begin{center}
\includegraphics[width=0.75\textwidth]{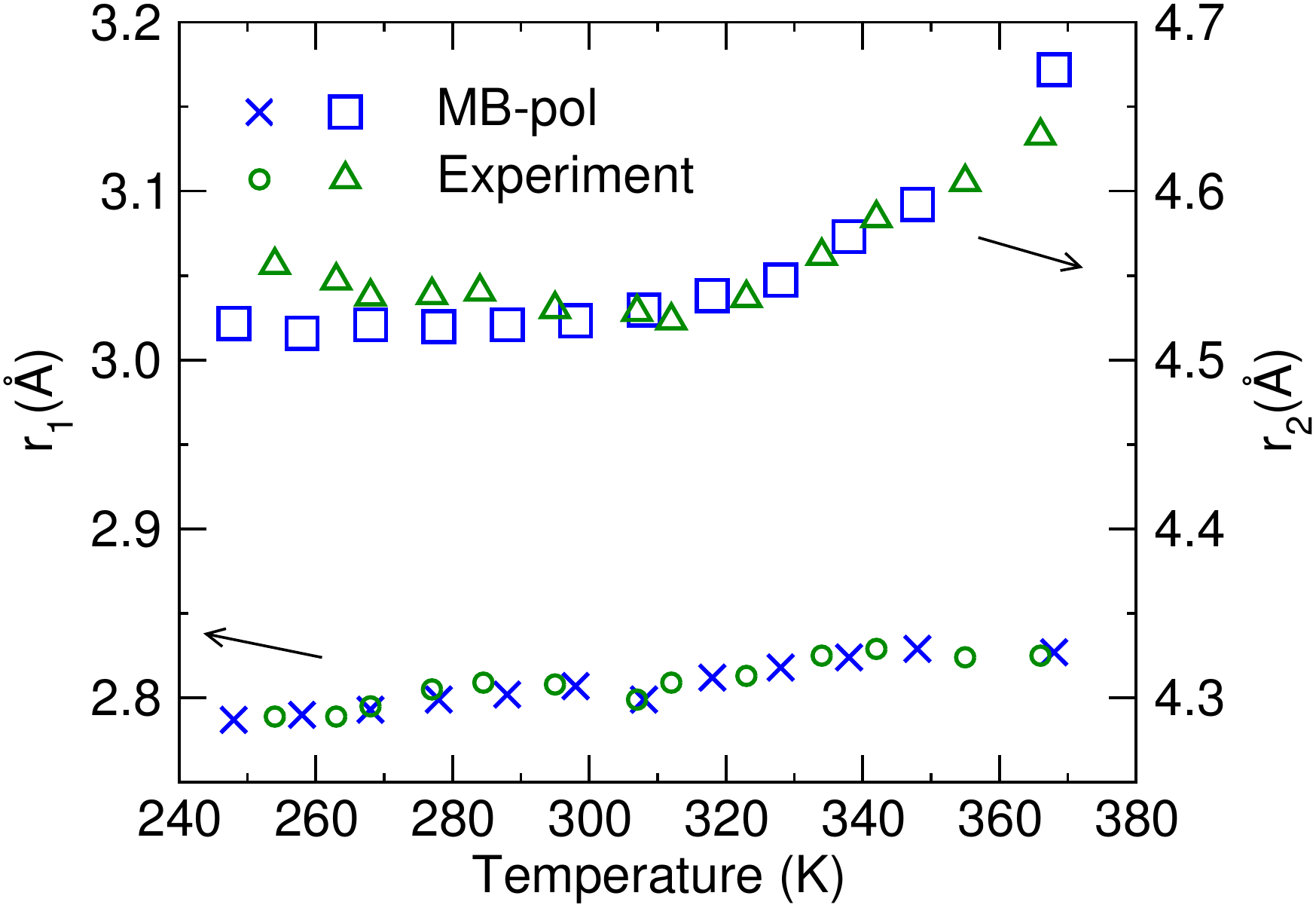}
\end{center}
\caption{Temperature dependence of the positions of the first and
second peaks of the oxygen-oxygen RDF of liquid water predicted 
by classical NPT simulations with MB-pol compared with the corresponding 
experimental data.\cite{SkinnerJCP2013,SkinnerJCP2014}
}
\label{fig:lpair_r1r2}
\end{figure}

Figure~\ref{fig:lpair_corr} shows the oxygen-oxygen (OO) radial distribution
function (RDF) calculated with MB-pol at several temperatures along with the
corresponding experimental curves derived from the
most recent X-ray diffraction measurements.\cite{SkinnerJCP2014,SkinnerJCP2013} 
The MB-pol RDFs are in good agreement with the experimental data at all temperatures,
although they slightly overestimate the height of the first peak. 
As discussed in previous studies, this difference, which increases with decreasing temperature, 
can be attributed to the the neglect of NQE and lead to more structural order in the
classical liquid than is experimentally observed.
It has been shown that, while the position of the first peak moves to larger r$_{OO}$ linearly with 
increasing temperature, the shift in the position of the second peak deviates from a 
linear dependence on r$_\textrm{OO}$ above $\sim$320~K.\cite{SkinnerJCP2014}
This trend is correctly captured by classical NPT simulations with MB-pol,
which reproduce the experimental data nearly quantitatively (Figure~\ref{fig:lpair_r1r2}).
The numerical comparison between MB-pol and experimental OO RDFs is
based on the position of the 1$^{st}$ (r$_1$) and 2$^{nd}$ 
(r$_2$) peaks, and the height of the 1$^{st}$ peak (g(r$_1$)) reported in
Table~\ref{tab:gr}. Altogether, these quantities are used to assess the
accuracy of MB-pol at 268, 278, and 308~K, through comparisons with 
the corresponding experimental values.\cite{SkinnerJCP2014,SkinnerJCP2013}

{\it Scoring.} 
In addition to the numerical comparison, a score was assigned to each of the three quantities
using Eq.~\ref{eq:lscore} with tolerance values of 0.1\%, 0.5\% and 5.0\%
respectively for the r$_1$, r$_2$ and g(r$_1$). The average
MB-pol scores obtained for the three quantities at 
at 268, 278, and 308~K are listed in Table~\ref{tab:gr}. 
The agreement between the calculated and experimental data 
is excellent at all temperatures, which
provides further evidence for the accuracy and transferability of MB-pol.

\begin{table*}[!htbp]
\caption{\label{tab:gr}
Positions of the first (r$_1$) and second (r$_2$) peaks (in \AA), and first peak height (g(r$_1$))
of the oxygen-oxygen RDF of liquid water predicted 
by classical NPT simulations with MB-pol at three different temperatures 
compared with the corresponding experimental data.\cite{SkinnerJCP2013,SkinnerJCP2014}
The temperatures at which the experimental measurements 
were performed are given in parenthesis.See text for details.
}
\begin{ruledtabular}
\begin{tabular}{ccccccccccccccccccc}
& \multicolumn{2}{c}{268 K} & \multicolumn{2}{c}{278 (277) K} & \multicolumn{2}{c}{308 (307) K} \\ 
	\cline{2-3} \cline{4-5} \cline{6-7} 
& MB-pol & Experiment & MB-pol & Experiment & MB-pol & Experiment  \\ 
\hline
r$_1$     & 2.796 & 2.798 & 2.806 & 2.797 & 2.818 & 2.806 \\
r$_2$     & 4.488 & 4.509 & 4.490 & 4.516 & 4.510 & 4.520 \\
g(r$_1$) & 3.044 & 2.801 & 2.936 & 2.723 & 2.678 & 2.484 \\
Average score & 90 & - & 93 & - & 93 & - \\
\end{tabular}
\end{ruledtabular}
\end{table*}

\subsection{Ice phases} 

\subsubsection{Melting point of hexagonal ice}
The melting point (T$_\textrm{m}$) of hexagonal ice (I$_\textrm{h}$) 
at the pressure of 1 atm was calculated
from classical MD simulations carried out in the NPT ensemble 
to characterize the ice/water coexistence as a function of temperature.
This approach, originally proposed in
Refs.\citenum{Woodcock1977,Woodcock1978a,Woodcock1978b}, has been used
to determine the melting point of several water models.
\cite{Karim1988,Karim1990,Bryk2002,Bryk2004,Wang2005,Vega2006}
The water and ice I$_\textrm{h}$ regions of the coexistence system, each containing 360
water molecules, were equilibrated independently at 300 and 100~K, respectively.
The two regions were then combined into a single
rectangular box of 720 water molecules (dimensions 22.57 $\times$ 23.46 $\times$
44.24 \AA$^{3}$) with the basal (0001) plane of ice I$_\textrm{h}$ in contact with the
liquid phase. Direct coexistence simulations were then carried out
in the NPT ensemble at different temperatures, during which the enthalpy of the combined
system was monitored. The melting point was determined as the
temperature at which the enthalpy of the combined system
remained constant throughout the entire trajectory.
The classical melting point of MB-pol is found at 263.5 $\pm$ 1.5 K.
The calculated T$_\textrm{m}$ is $\sim$9~K lower than the experimental value and in good agreement with
the classical MB-pol estimate of 263~K for the temperature of maximum density (see
Section~\ref{subsection:density}).

\subsubsection{Lattice energies and densities of ice phases} 
Table \ref{ice_ener_dens} lists the lattice energies of different ice phases
predicted by MB-pol in comparison with the corresponding experimental estimates.  
The MB-pol values were obtained from single point calculations 
carried out for the experimental crystal structures taken from Ref.
\citenum{PhysRevLett.107.185701}. In all cases, the MB-pol results are in excellent
agreement with the experimental values. The maximum deviation of 
$\sim$2\% is found for ice VIII, a high density phase. 
A detailed analysis of the lattice energies in terms of many-body 
contributions will be reported in a forthcoming publication. 

\begin{table}[!ht]
\centering
\caption{
	Melting point (in K) of hexagonal ice along with lattice energies (in meV) and densities (in g cm$^{-3}$) of several ice phases calculated using MB-pol.
	Scores assigned to each computed quantity are also listed. See text for details.
}
\label{ice_ener_dens}
\begin{tabular}{lccccccccccccccc}
\hline\hline
\multirow{2}{*}{Ice} & \multicolumn{3}{c}{Melting point} & & \multicolumn{3}{c}{Lattice energy }  &  & 
 &  & 
\multicolumn{2}{c}{Density } & 
 &   \\ \cline{2-4} \cline{6-8}\cline{10-14}
 &  MB-pol & Exp.  & Score  &    
&  MB-pol & Exp.\footnote{From Ref. \citenum{whalley1984energies}}  & Score  &   & 
  T(K) & P(bar) & MB-pol & Exp. &  Score    \\
\hline\hline
I$_\textrm{h}$ & 263.5  & 273 & 90 & & -611     &  -610   &  100  &  &   250 & 0     & 0.921   & 0.920\footnote{From Ref. \citenum{Rottger:sh0050}}    &       100 \\
II                     &  -         &  -     & -    & & -603     &  -609   &  100  &  &   123 & 0     & 1.198   & 1.190\footnote{From Ref. \citenum{Fortes:aj5041} }    &       100 \\
VIII                 &  -          &  -    & -     & & -590     &  -577   &  80   &  &   263 & 24000 & 1.658   & 1.611\footnote{From Ref. \citenum{kuhs1984structure}} &        80 \\
IX                   &  -          &  -    & -     & & -601     &  -606   &  100  &  &   165 & 2800  & 1.208   & 1.194\footnote{From Ref. \citenum{londono1993neutron} } &        90 \\
XIII                 &  -          &  -    & -     & & -595     &    -       &   -      &  &   80  & 1     & 1.281   & 1.251\footnote{From Ref. \citenum{petrenko1999physics} } &     80 \\
XIV                 &  -         &  -     & -     & & -592     &    -      &   -      &  &   80  & 1     & 1.312   & 1.294\footnote{From Ref. \citenum{petrenko1999physics} }  &     90 \\
XV                  &  -         &  -     & -     & & -587     &    -      &   -      &  &   80  & 0     & 1.364   & 1.326\footnote{From Ref. \citenum{PhysRevLett.103.105701} } &    80 \\
Average score &           &         & 90  &  &            &           &    95  &  &         &        &             &                                                                                                       &    89 \\
\hline\hline
\end{tabular}
\end{table}

The densities of several ice phases calculated from classical NPT simulations with MB-pol 
at different temperatures and pressures are compared in Table~\ref{ice_ener_dens} with the
corresponding experimental data.
\cite{Rottger:sh0050,Fortes:aj5041,kuhs1984structure,londono1993neutron,
petrenko1999physics,petrenko1999physics,PhysRevLett.103.105701} 
Excellent agreement is found between the theoretical predictions 
and the experimental values, with the largest relative
error being less than 3\% for ice VIII.  
It should be noted that the classical NPT simulations with MB-pol slightly
overestimates the densities of all ice phases, which again indicates 
that explicit inclusion of NQE is necessary for more quantitative
comparisons with the experimental data.
In this context, it was demonstrated that quantum simulations are indeed
strictly required to correctly compare the theoretical results with the
experimental data at temperatures below 100 K.\cite{mcbride2009quantum} 
Similar to liquid water,\cite{Francesco1JCTC2014}  
NQE are expected to lower the density and, therefore, 
improve the agreement with experiment, especially for the ice phases (XIII, XIV, and XV)
that are stable at lower temperature.

{\it Scoring.} 
Adopting the same scoring scheme used for the liquid properties (Eq.~\ref{eq:lscore}), 
scores were assigned to both lattice energies and densities calculated with MB-pol
for the different ice phases considered in this analysis, setting the tolerance value to 1\% for both properties. For
the melting point, it is set to 2.5\% which is the same as used in Ref.~\citenum{VegaPCCP2011}.
As shown in Table~\ref{ice_ener_dens}, the average scores 
received by MB-pol are 95 and 89 for ice lattice energies and densities, respectively,
which, combined with the analyses reported in the previous sections, demonstrate that 
MB-pol consistently achieves high accuracy in predicting the properties of water
from clusters to the liquid phase and ice.

\section{Conclusions} 
In this study, the accuracy of the MB-pol many-body potential 
is assessed from extensive analysis of the energetics
as well as of spectroscopic, structural, thermodynamic,
and dynamical properties of water from the gas to the condensed phase.

The analysis of gas-phase properties shows that 
both individual many-body contributions to the interaction energies and 
harmonic frequencies calculated for water clusters up to the hexamer with MB-pol are in
excellent agreement with reference data obtained at the
coupled cluster level of theory. The largest deviations are
observed for cyclic clusters due to the accumulation of errors associated with
repeating, symmetry-equivalent two-body and three-body units in the MB-pol
representation of the underlying Born-Oppenheimer PES. 

For the liquid phase, classical MD simulations carried out with MB-pol
correctly reproduce the temperature dependence of all structural, thermodynamic,
and dynamical properties analyzed in this study. The deviations from the experimental values increase as the
temperature decreases. Since MB-pol was derived entirely from electronic structure data and
thus represents the water Born-Oppenheimer PES, 
the differences between results obtained from classical MD simulations with MB-pol and experimental data 
are expected and confirm that nuclear quantum effects must be explicitly taken into
account in the simulations with MB-pol for a quantitative (and rigorous) 
description of the molecular properties of liquid water.
Finally, the melting point of hexagonal ice as well as 
both lattice energies and densities calculated with MB-pol for several ice phases 
are found in good agreement with the corresponding experimental data, which 
provides further evidence for the transferability of MB-pol. 

Besides demonstrating the high and, in many respects, unprecedented accuracy 
with which MB-pol predicts the properties of water across different phases,
this study also provides a series of rigorous tests that should be
used to assess the ability of both empirical and \textit{ab initio}
water models ``to get the right answers for the right reasons",\cite{FrancescoACR2016} which
is the fundamental prerequisite for a physically correct understanding
of the behavior of water at the molecular level.

\section{Supplementary Material}
List of 
Additional data related to the analysis of many-body interactions and vibrational frequencies of water clusters
along with the corresponding molecular coordinates. Tables with numerical values for all water properties shown
in Figures 4 - 11.

\section{Acknowledgements}
This research was supported by the
National Science Foundation through Grant CHE-1453204. 
This work was performed in part under
the auspices of the US DOE by LLNL under Contract DE-AC52-07NA27344.
This research used resources of the Argonne Leadership Computing Facility, 
which is a DOE Office of Science User Facility supported under Contract DE-AC02-06CH11357,
as well as the Extreme Science
and Engineering Discovery Environment (XSEDE), which is
supported by the National Science Foundation (Grant ACI-1053575). 
S.S. acknowledges the University of California, San Diego for financial support through the 
Frontiers of Innovation Scholars Program (FISP).

\vspace{1.cm}

The submitted manuscript has been created by UChicago Argonne, LLC, 
Operator of Argonne National Laboratory (``Argonne"). Argonne, 
a U.S. Department of Energy Office of Science laboratory, 
is operated under Contract No. DE-AC02-06CH11357. 
The U.S. Government retains for itself, 
and others acting on its behalf, a paid-up nonexclusive, 
irrevocable worldwide license in said article to reproduce, 
prepare derivative works, distribute copies to the public, 
and perform publicly and display publicly, by or on behalf of the Government. 
\clearpage 

\bibliography{references}

\end{document}